%% file: main.tex
\documentclass[letterpaper]{article} 
\usepackage[preprint]{aaai2027}  
\usepackage[hyphens]{url}  
\usepackage{graphicx} 
\usepackage{natbib}  
\usepackage{caption} 
\usepackage{booktabs}

\usepackage{amsmath}
\usepackage{amssymb}

\usepackage{tabularx}
\usepackage{array}
\usepackage{longtable}
\usepackage{pdflscape}
\usepackage{seqsplit}

\title{Pre-Registered External Evaluation Yields a Consistent Partial-Replication Category across Three Transcriptomic Foundation Models}

\author{
    Mehrdad Shoeibi,
    Niloofar Yousefi
}
\affiliations{
    University of Central Florida
}

\begin{document}

\maketitle

\begin{abstract}
Transcriptomic foundation models are increasingly used as reusable cell and
gene representations, but validating them on new data under weak supervision
and distribution shift is hard: standard comparisons conflate genuine
representation signal with model capacity, row-identity artifacts, gains over
strong task-specific baselines, and outcome rules chosen after seeing the test
set. We introduce a pre-registered, final-test-once evaluation framework that
locks the outcome rule, seeds, and target-gene-grouped splits before any test
data are seen, and scores each frozen representation against a strong
expression baseline, a matched-capacity Gaussian control, and a within-split
row-identity (shuffle) control; only the per-cell embedding-extraction step is
model-specific. Applying it to three architecturally
distinct models---Geneformer, scGPT, and UCE---across two external Replogle
Perturb-seq datasets (RPE1 and K562), all three clear the capacity and
row-identity controls by a wide margin, yet none reliably beats the expression
baseline: the strongest (Geneformer) exceeds it by at most about $0.03$ test
$R^2$ and clears the pre-registered four-of-five-seed threshold in neither
dataset, while scGPT and UCE fall below it. All three therefore land in the
same pre-registered partial-replication category---a consistent
cross-architecture outcome, even though the baseline-relative gap differs in
sign and magnitude across models. These representations carry real structure
beyond trivial controls but, under this weak magnitude label, do not transfer
past a simple strong baseline; the locked framework is reusable for any frozen
transcriptomic representation by swapping only the extraction step.
\end{abstract}

\input{sections/01_introduction}
\input{sections/01b_related_work}
\input{sections/02_methods}
\input{sections/03_results}
\input{sections/04_discussion}
\input{sections/05_limitations}

\bibliography{references}

%
\onecolumn
\let\origunderscore\_
\renewcommand{\_}{\origunderscore\allowbreak}
\sloppy
\emergencystretch=3em

\input{appendix/appendix}

\end{document}

%% file: sections/01_introduction.tex
\section{Introduction}

\subsection{Motivation: evaluating transcriptomic foundation models under weak supervision and distribution shift}

Transcriptomic foundation models increasingly provide reusable cell and
gene representations for downstream prediction in perturbation biology
and single-cell analysis. Here labels are often weak or proxy-based
(constructed from distributional comparisons against reference
populations rather than mechanistic phenotypes), and data are generated
in one experimental context and evaluated on different populations, so
the representation, the weak target, the domain baseline, and the
evaluation protocol can interact in ways that make transfer claims hard
to disentangle. Before such representations can be called transferable
under perturbation shift, they need external-validation protocols that
separate representation signal from model capacity, row-identity
artifacts, strong domain baselines, and post-hoc outcome interpretation.
This work addresses that problem; its contribution is the evaluation
framework itself, not a new model. Because both the weak label and the
predictors are derived from the same underlying transcriptomic
measurements, the study evaluates representation efficiency for an
expression-derived proxy, not prediction of an independently measured
phenotype.

\subsection{Why current evaluation is insufficient}

Simple performance comparisons cannot establish that a representation
transfers: a frozen representation can beat random and shuffled controls
without uniformly improving over a strong task-specific baseline, and
apparent gains can arise from random feature capacity, within-split
row-identity artifacts, weak-label idiosyncrasies, or retrospective
threshold interpretation after the test set is observed. Credible
external validation therefore needs matched-capacity and row-identity
controls, a strong domain baseline, target-grouped splits, fixed seeds,
thresholds locked before the test data are seen, and a final-test-once
protocol. Evaluation should ask three separable questions: whether a
representation carries non-random signal, whether that signal exceeds a
strong domain baseline, and whether the gain is stable across seeds
under a pre-specified rule.

\subsection{Contributions}

\begin{itemize}
\setlength{\itemsep}{1pt}\setlength{\parskip}{0pt}
\item \textbf{A pre-registered, final-test-once external-validation framework} for frozen transcriptomic foundation-model representations under weak supervision and distribution shift, comprising a locked outcome cascade, a strong expression baseline, a matched-capacity Gaussian control, a within-split Shuffle control, \texttt{target\_gene}-grouped train/validation/test splits, five fixed training seeds, machine-epsilon validation-parity checks, SHA-256 fingerprints on the locked label and split assignments, and a reviewer-facing reproducibility audit trail.
\item \textbf{A three-architecture application} to frozen representations from Geneformer \citep{theodoris2023geneformer}, scGPT \citep{cui2024scgpt}, and UCE \citep{rosen2023uce}, across two external Replogle Perturb-seq phases \citep{replogle2022perturbseq} (RPE1 and K562 GWPS), under a fixed Anderson-Darling weak label, the locked five-model family (expression baseline, foundation-model features, dimension-matched Gaussian control, within-split Shuffle control, and MAG-only anchor), and the locked decision cascade.
\item \textbf{A multi-architecture boundary-finding result.} All three representations close as \texttt{PARTIAL\_EXTERNAL\_REPLICATION} in both phases, producing a common \texttt{PARTIAL\_PARTIAL} cross-dataset pattern: each exceeds the Gaussian and Shuffle controls, yet none meets the locked STRONG criterion against the strong expression baseline. The shared category is a consistency result across architectures, not evidence that the protocol separates the models; at the criterion level the baseline-relative gains differ in sign and magnitude.
\end{itemize}

\subsection{Preview of result and scope}

The central claim is deliberately bounded. Under a locked
final-test-once protocol, Geneformer, scGPT, and UCE each show a
\texttt{PARTIAL\_PARTIAL} external-validation pattern across two external
Replogle Perturb-seq phases (RPE1 and K562 GWPS), with consistent
evidence against random-capacity and row-identity explanations but
without satisfying the locked STRONG criterion against a strong
expression baseline. The result locates a transfer boundary that holds
across three representation families under weak supervision and
distribution shift, and the framework that exposes it is reusable for
any frozen transcriptomic representation by swapping the per-cell
extraction step while the downstream evaluation layer is held fixed.

%% file: sections/01b_related_work.tex
\section{Related Work}

\subsection{Transcriptomic foundation models}

A growing family of single-cell transcriptomic foundation models has been
proposed for general-purpose representation learning, pretrained on tens
of millions of profiles. Geneformer \citep{theodoris2023geneformer} uses
rank-based gene tokenization and a transformer backbone; scGPT
\citep{cui2024scgpt} adapts a generative pretrained transformer to
single-cell multi-omics; scFoundation \citep{hao2024scfoundation} scales
to roughly 100M parameters over 50M profiles; and Universal Cell
Embeddings \citep{rosen2023uce} target a cross-species, cross-tissue
latent space without fine-tuning. Our study evaluates three of these
models, Geneformer, scGPT, and UCE, under a single representation-agnostic
framework in which only the per-cell embedding-extraction step is
model-specific.

\subsection{Perturbation-prediction methods}

A separate line of work develops task-specific predictors for in-silico
perturbation effects. GEARS \citep{roohani2024gears} combines a deep
network with a Gene Ontology knowledge graph to predict transcriptional
outcomes of single and multigene perturbations. GPerturb
\citep{xing2025gperturb} introduces a Gaussian-process sparse perturbation
regression that estimates gene-level effects with uncertainty quantification.
The present work does not propose a new perturbation predictor; instead,
we evaluate frozen foundation-model representations against task-specific
expression baselines and locked controls under a pre-registered decision
rule. Predictors such as GEARS and GPerturb could be substituted into the
present framework as task-specific comparators in future studies.

\subsection{Benchmarking of foundation models on perturbation tasks}

Several benchmarks find that single-cell foundation models do not
reliably beat simple baselines. \citet{ahlmanneltze2025linear} compared
five foundation models (Geneformer, scGPT, scFoundation, scBERT, UCE)
with GEARS and CPA against simple additive and mean-prediction baselines
on Norman, Adamson, and Replogle data and found none consistently better.
\citet{bendidi2024pca} reported that PCA and scVI remained competitive or
superior to foundation-model embeddings on Replogle and L1000 data, with
random embeddings approaching them on batch-effect tasks.
\citet{kedzierska2025zeroshot} found Geneformer and scGPT inconsistent
relative to highly variable gene selection, scVI, and Harmony on
zero-shot clustering and reconstruction. \citet{csendes2025benchmark}
found a training-mean baseline competitive or superior to scGPT and
scFoundation across Adamson, Norman, and both Replogle datasets, and
\citet{wenteler2025perteval} reported limited improvement over baselines
under distribution shift with PertEval-scFM. Beyond perturbation tasks,
\citet{boiarsky2024deeper} found an L1-regularized logistic-regression
baseline competitive with scBERT and scGPT on cell-type annotation, and
\citet{wu2025biology} benchmarked six foundation models (including scGPT,
UCE, scFoundation, and Geneformer) and reported task-dependent
performance with no single model dominating the baselines.

A concurrent submission by an overlapping author set
\citep{anon2026magnitude} studies the Virtual Cell Challenge benchmark and
identifies response magnitude as the dominant low-dimensional signal for
held-out target genes. That work analyses which signal drives predictive
performance on a single benchmark; the present paper instead pre-registers a
locked decision rule and applies it final-test-once to three frozen
representations across two external Replogle datasets. The two are therefore
distinct in question, protocol, and evaluated datasets.

The present work is empirically aligned with these findings: in our
locked external-validation study, frozen Geneformer-, scGPT-, and
UCE-derived features each carry non-random transfer signal (clearly
exceeding matched-capacity and within-split row-identity controls) yet
do not consistently surpass a strong expression baseline, and the locked
STRONG criterion is met by none of them in either external phase. The
methodological contribution differs from these benchmarks in four
respects. First, the protocol, controls, seeds, splits, and thresholds
are pre-registered and locked before the final-test data are evaluated.
Second, two locked controls (a matched-capacity Gaussian and a
within-split Shuffle) are integrated into a priority cascade that
separates random capacity, row-identity artifacts, and gains over a
strong baseline. Third, the framework yields an explicit boundary-finding
outcome (\texttt{PARTIAL\_EXTERNAL\_REPLICATION}) under a final-test-once
protocol rather than a positive-versus-negative judgment. Fourth, it is
representation-agnostic by construction: only the extraction step is
model-specific, so the same locked protocol applies to any frozen
transcriptomic representation by swapping that step alone.

\subsection{Pre-registration and external validation in biomedical ML}

The need for pre-registration combined with external validation in
biomedical predictive modeling has been articulated in adjacent domains.
\citet{gallitto2025external} proposed registered-model designs with
adaptive sample splitting for biomedical ML, advocating public deposition
of feature-processing steps and model weights before external validation.
We are not aware of prior Perturb-seq foundation-model evaluations that
jointly combine pre-registered, final-test-once external validation, a
matched-capacity Gaussian control, a within-split row-identity control,
machine-epsilon validation-parity safeguards, and SHA-256 fingerprints
on labels and split assignments.

%% file: sections/02_methods.tex
\section{Methods}

\subsection{Study design and pre-registration}

\begin{figure*}[t]
\centering
\includegraphics[width=0.82\textwidth]{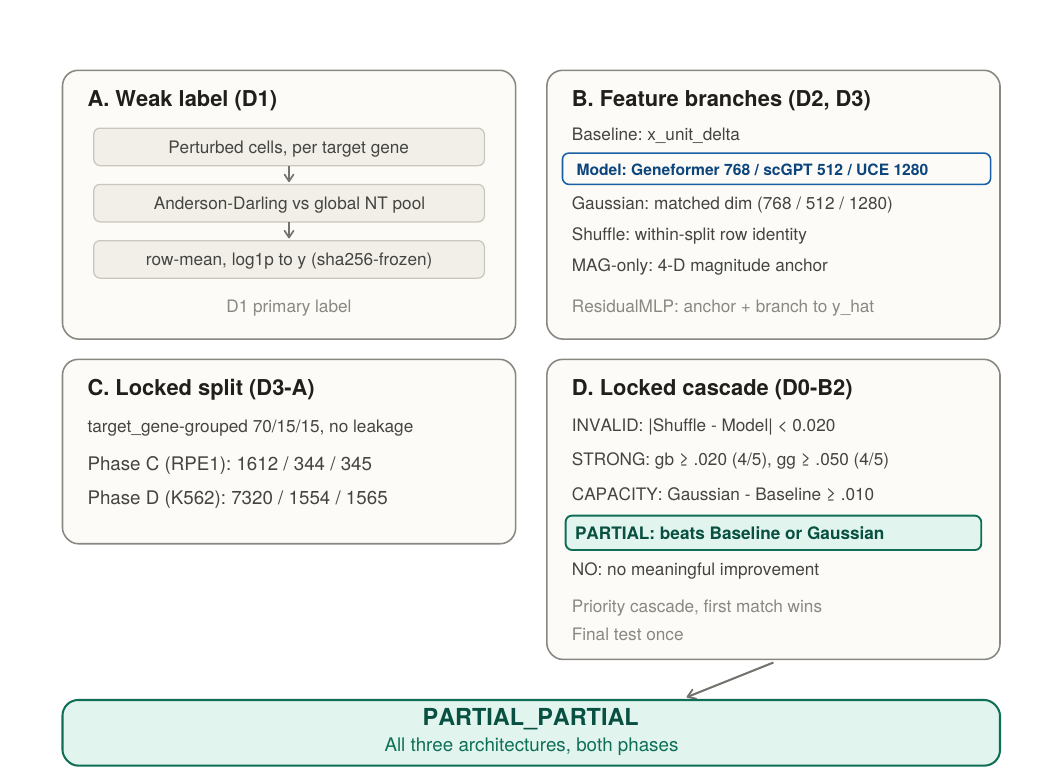}
\caption{Locked external-validation workflow. Four-panel design schematic of the
pre-registered, final-test-once framework: (A) weak-label construction, (B) the
five-model feature family into which each frozen representation plugs
(Geneformer, scGPT, and UCE, with embedding dimensions 768, 512, and 1280), (C)
the locked \texttt{target\_gene}-grouped split, and (D) the locked decision-rule
cascade with thresholds. All three representations close as
\texttt{PARTIAL\_EXTERNAL\_REPLICATION} in both phases. Panel~D is
schematic; the exact outcome rule is given in Methods~\S\ref{sec:cascade}.}
\label{fig:f1_design_schematic}
\end{figure*}

This is a pre-registered external-validation study of transcriptomic
foundation-model transfer under weak supervision and distribution shift.
The weak label, expression baseline, controls,
\texttt{target\_gene}-grouped splits, seeds, decision cascade, and
thresholds were locked in a pre-registration manifest with the Geneformer
arm, before its final-test evaluation (provenance in Appendix
Table~\ref{tab:app_t3_artifact_lineage}). The scGPT and UCE arms are
protocol-preserving extensions: each has its per-model extraction rule
committed before its own final-test evaluation, reuses the already-locked
protocol without modification, and is evaluated final-test-once on the
same held-out partition. We therefore do not claim an independent
pre-registration for scGPT and UCE, only that they were evaluated under
the previously locked protocol. Each phase, Phase C (RPE1) and Phase D (K562 GWPS), ran
through a pre-flight design lock, a train/validation sanity stage
verifying determinism, and a single final-test evaluation
(\S\ref{sec:finaltest}). Figure~\ref{fig:f1_design_schematic} summarizes
the workflow; the contribution is the locked evaluation framework itself,
not a new model architecture.

\subsection{External datasets and analysis units}

Two external CRISPRi Perturb-seq datasets are used
\citep{replogle2022perturbseq}: Phase C uses the Replogle 2022 RPE1
genome-wide dataset and Phase D the Replogle 2022 K562 genome-wide
dataset. The primary row unit is the \texttt{gene\_transcript}, a
perturbation-and-transcript-resolved unit distinct from
\texttt{target\_gene}. All splits are constructed at the \texttt{target\_gene}
level, so a single target gene cannot appear in more than one of the
train, validation, or test partitions; without this grouping the same
perturbation would occur in both training and test and inflate apparent
generalization. Eligible-row counts (10{,}439 in Phase D, 2{,}301 in Phase C), the
deterministic 70/15/15 split seeds, and the partition sizes are
summarized in Appendix Table~\ref{tab:t1_dataset_split}; per-split
unique-gene counts and split-integrity checks (zero \texttt{target\_gene}
leakage) are in Appendix Table~\ref{tab:app_t5_repro_checklist}.

\subsection{Weak-label construction}

Ground-truth perturbation-effect labels are unavailable, so we use a
per-row response-magnitude weak label based on the k-sample
Anderson-Darling (AD) distance \citep{scholz1987ksample} between the
perturbed-cell distribution for a \texttt{gene\_transcript} and a global
non-targeting (NT) reference (75{,}328 NT cells in Phase D):

\begin{quote}
y = log1p(max(0, mean finite per-feature-gene Anderson-Darling
statistic on raw counts vs global NT)).
\end{quote}

Both phases use the identical label form (the Phase D manifest attests
that the Phase C logic was mirrored exactly; \texttt{scipy} version and
\texttt{anderson\_ksamp} settings are pinned in the Appendix). The label
captures the magnitude of the distributional shift a perturbation
induces, not a mechanistic phenotype. A secondary per-gem\_group
diagnostic label (D1.5) was constructed for transparency only and is
excluded from outcome assignment (\texttt{d15\_used\_for\_outcome = false}
in every Phase D QC).

\subsection{Feature representations}
\label{sec:features}

For each row, expression features use a fixed preprocessing recipe:
per-cell counts-per-10{,}000 (CP10K) normalization followed by
\texttt{log1p}, averaged separately over the row's perturbed cells and the
global NT pool, giving the expression delta
\texttt{x\_delta = mean(log1p(CP10K))\_pert - mean(log1p(CP10K))\_NT}
(8{,}248-D in Phase D; 8{,}749-D in Phase C). Its L2-unit-normalized
form \texttt{x\_unit\_delta} supplies the strong expression baseline branch.
A four-dimensional magnitude vector MAG (mean, scaled L2 norm, max, and
std of \texttt{x\_delta}) feeds a non-trivial anchor.

Foundation-model representations are frozen (no fine-tuning); the
downstream evaluation specification (label, splits, seeds, control
definitions, model family, and decision cascade) is held fixed across
arms, while the per-cell embedding-extraction step is model-specific and
isolated upstream. Each arm is executed in a separately committed chain,
so downstream numerical outputs are not assumed byte-identical across arms
(Appendix, \emph{Extended limitations}). Geneformer-V2-104M (gc104M vocabulary,
768-D) tokenizes raw counts with a TranscriptomeTokenizer and extracts
per-cell embeddings with EmbExtractor (CLS mean-pool, layer $-1$, fp32).
scGPT (512-D, \texttt{whole\_human} checkpoint) and UCE (1280-D, 33-layer
checkpoint) produce embeddings from their own published frozen
checkpoints; the per-row aggregation
(\texttt{z\_emb = mean(emb)\_pert - mean(emb)\_NT}, then L2-unit-normalized)
is identical across arms. Model-specific tokenization, vocabulary and
gene matching, normalization and binning, extraction layer, pooling,
precision, and sequence handling are \emph{not} harmonized; the exact
scGPT and UCE configurations, read from the committed extraction scripts,
are reported in the Appendix (\emph{scGPT and UCE extraction
specifications}), so cross-arm differences reflect the encoder together
with its preprocessing, not the encoder alone. Two locked
controls match each arm's capacity and use no labels: a Gaussian
random-feature control
\texttt{g\_all(s) = unit\_norm(RandomState(s+777).randn(n, d))} of
dimension \texttt{d} matching the representation, and a within-split
Shuffle control (a within-split feature--row correspondence control)
permuting the representation within each split; both are deterministic
from the frozen seeds.

\subsection{Models and training protocol}

A single locked supervised family is used. MAG features feed a
\texttt{RidgeCV}-style anchor whose $\ell_2$ penalty is selected by
train-only \texttt{GroupKFold(5)} over a locked \texttt{ALPHA\_GRID}
(anchor \texttt{$\alpha$ = 0.01} in both phases and identical across arms;
the separately fitted \texttt{Baseline} branch instead selected slightly
different penalties across arms in Phase C, Appendix \emph{Extended
limitations}). Each representation then
predicts the residual \texttt{r = y - anchor(MAG)} with an additive
\texttt{ResidualMLP} branch (hidden layers \texttt{{[}256, 128{]}}, ReLU,
dropout 0.1; Adam, learning rate 1e-3, weight decay 1e-4, batch size 64;
MSE objective; fixed 25 epochs, no early stopping, CPU). \texttt{MAG-only}
is the anchor prediction alone. Every \texttt{(model, seed)} pair over the
five locked seeds \texttt{{[}20260527, 20260528, 20260529, 20260530,
20260531{]}} is trained independently, with no test-set tuning, no
per-test sweep, and no post-hoc seed selection.

\subsection{Evaluation metrics}

The primary metric is test-partition $R^2$; Spearman $\rho$ and mean
absolute error (MAE) are secondary. Each metric is computed once per
\texttt{(model, seed)} pair, then reported as mean $\pm$ std (sample
standard deviation, \texttt{ddof=1}) across the five seeds. Paired test-$R^2$ deltas (for example \texttt{Model - Baseline})
are computed per seed and averaged, so a reported phase-mean delta is
the mean over the five seeds of per-seed test-$R^2$ deltas, not a delta
of prediction-pooled $R^2$; these feed the locked cascade. No p-values
are used, and the supplementary bootstrap intervals (\S\ref{sec:supp})
are not significance tests.

\subsection{Locked decision-rule cascade}
\label{sec:cascade}

The outcome for each phase is assigned by a locked priority cascade
with thresholds pre-registered in the D0-B2 manifest. The cascade is
defined over a generic representation arm: \texttt{Model} denotes that
arm's frozen representation (Geneformer, scGPT, or UCE), evaluated
against that arm's own \texttt{Baseline}, \texttt{Gaussian}, and
\texttt{Shuffle} branches. The priority order is

\begin{quote}
\texttt{INVALID > STRONG\_EXTERNAL\_REPLICATION > CAPACITY\_ONLY\_OR\_AMBIGUOUS > PARTIAL\_EXTERNAL\_REPLICATION > NO\_EXTERNAL\_REPLICATION}.
\end{quote}

\textbf{INVALID} triggers if \texttt{|Shuffle - Model| < 0.020} at the
phase-mean test-$R^2$ level. \textbf{STRONG} requires all four locked
criteria: \texttt{Model - Baseline} mean $\geq$ +0.020 \emph{and}
$\geq$4/5 seeds $\geq$ +0.020, \emph{and} \texttt{Model - Gaussian} mean
$\geq$ +0.050 \emph{and} $\geq$4/5 seeds $\geq$ +0.050.
\textbf{CAPACITY\_ONLY\_OR\_AMBIGUOUS} triggers if
\texttt{Gaussian - Baseline} mean $\geq$ +0.010 (a random-capacity branch
already beating the baseline). \textbf{PARTIAL} is assigned when none of
the above trigger and at least one of the baseline-relative or
Gaussian-relative mean deltas is positive without shuffle invalidation;
\textbf{NO} otherwise. The thresholds (+0.020, +0.050, +0.010, 0.020)
are decision-rule criteria, not statistical-significance criteria. The
manifest was pre-registered with the Geneformer arm (in which
\texttt{Model} is Geneformer) and applied unchanged to the scGPT and UCE
arms.

\subsection{Final-test-once protocol and leakage controls}
\label{sec:finaltest}

The held-out partition is evaluated exactly once per phase, after the
train/validation sanity stage. Validation parity
between the final-test and sanity pipelines holds at maximum
$|\Delta| = 1.110\mathrm{e}{-16}$ across all twenty-five
\texttt{(model, seed)} validation-metric pairs (per-phase, per-metric
validation-parity summaries in Appendix
Table~\ref{tab:app_t5_repro_checklist}), providing
numerical evidence that the final-test and sanity pipelines were
functionally identical to machine precision on the validation partition. The \texttt{target\_gene}-grouped
split yields zero cross-split target-gene leakage in both phases, and
the Phase D primary-label and split-assignment SHA-256 fingerprints
were re-verified at the start of the final test (values in the
Appendix, \emph{Artifact fingerprints and manifest details}). The test
split was never reused for model selection.

\subsection{Supplementary analyses, reproducibility, and artifact trail}
\label{sec:supp}\label{sec:artifact-trail}

After the locked outcomes were assigned, four analyses were computed on
the committed frozen test predictions: (A1) paired cluster-bootstrap
95\% confidence intervals on test-$R^2$ deltas (\texttt{B = 2000}; here
each cluster is a single test row, so it reduces to a paired row-level
bootstrap; full specification in the Appendix); (A2) a paired seed-level
threshold-robustness summary; (A3) a cross-dataset consistency matrix;
and (A4) a baseline-competitiveness audit. These operate on frozen
predictions, do not re-run the final test, and do not enter the locked
cascade. \emph{Interpretive boundary:} these intervals do not replace the
locked cascade, do not define statistical significance against the
+0.020 / +0.050 thresholds, and do not alter the outcomes. The committed
artifact lineage (Appendix Table~\ref{tab:app_t3_artifact_lineage}), all
seeds and SHA-256 fingerprints, and a reviewer-facing reproducibility
checklist (Appendix Table~\ref{tab:app_t5_repro_checklist}) are provided
in the Appendix.

%% file: sections/03_results.tex
\section{Results}

This section reports the frozen final-test outcomes for both external
phases under the locked decision-rule cascade (Methods
\S\ref{sec:cascade}) and the supplementary frozen-output analyses
(Methods \S\ref{sec:supp}). All numerical values come from the locked
final-test and supplementary analyses. The locked outcomes (Phase C and
Phase D both \texttt{PARTIAL\_EXTERNAL\_REPLICATION}; cross-dataset pattern
\texttt{PARTIAL\_PARTIAL}) are not modified by this section.

\subsection{Locked final-test performance across two external datasets}

Under the locked, final-test-once protocol, D-Geneformer attained the
highest mean test $R^2$ in both phases. In Phase C (RPE1; n = 345 test
rows) the mean test $R^2$ values were
\texttt{C-MAG-only = 0.7269}, \texttt{C-Baseline = 0.8310},
\texttt{C-Geneformer = 0.8362}, \texttt{C-Gaussian = 0.6587}, and
\texttt{C-Shuffle = 0.6441}. In Phase D (K562 GWPS; n = 1{,}565) they were
\texttt{D-MAG-only = 0.5778}, \texttt{D-Baseline = 0.7652},
\texttt{D-Geneformer = 0.7944}, \texttt{D-Gaussian = 0.5422}, and
\texttt{D-Shuffle = 0.5315}. Spearman $\rho$ and MAE rank the families in
the same order in both phases (full metrics in Appendix
Tables~\ref{tab:t2_final_performance},~\ref{tab:app_t1_per_seed_metrics}
and Appendix Figure~\ref{fig:f2_test_performance}). In both phases the
expression baseline ranked second by mean test $R^2$, a high-performing
in-framework comparator rather than a weak strawman
(\S\ref{sec:baseline-competitiveness}).

\subsection{Locked decision-rule outcomes}

Applied exactly once to each phase's final-test partition, the cascade
assigned \texttt{PARTIAL\_EXTERNAL\_REPLICATION} in both. STRONG was not
assigned because the locked \texttt{GF - Baseline} seed-consistency
criterion ($\geq$4/5 seeds with per-seed delta $\geq$ +0.020) failed in
both phases (Phase C: 1/5; Phase D: 3/5), even though the
\texttt{GF - Gaussian} seed-consistency criterion passed at 5/5 in both.
The capacity-only trigger (\texttt{Gaussian - Baseline} mean $\geq$ +0.010)
was inactive (Phase C \texttt{-0.1724}; Phase D \texttt{-0.2230}), and the
invalidity trigger (\texttt{|Shuffle - Geneformer| < 0.020}) was inactive
(Phase C \texttt{0.1920}; Phase D \texttt{0.2629}). Appendix
Table~\ref{tab:t3_decision_rule} and Appendix
Figure~\ref{fig:f3_decision_rule_deltas} give the full
criterion-by-criterion view.

\subsection{Control analyses disfavor the specified capacity-only and shuffle-artifact explanations}
\label{sec:control}

The matched-capacity Gaussian and within-split Shuffle controls
collapsed well below both Geneformer and Baseline in both phases (values
in Appendix Table~\ref{tab:t2_final_performance}). Geneformer clears the
matched-capacity control, \texttt{GF - Gaussian = +0.1775} (C) and
\texttt{+0.2521} (D), at the \texttt{+0.050} Gaussian criterion (5/5
seeds), and lies far from the within-split control,
\texttt{|Shuffle - Geneformer| = 0.1920} (C) and \texttt{0.2629} (D), above
the \texttt{0.020} invalidity threshold it gates; \texttt{Gaussian -
Baseline} is negative in both phases (\texttt{-0.1724},
\texttt{-0.2230}). These
provide evidence against the specified random-capacity and within-split
row-identity explanations under their two constructions; they do not
exclude every possible capacity, leakage, shared-preprocessing, or
gene-frequency artifact (Appendix
Table~\ref{tab:t3_decision_rule}, Appendix
Figure~\ref{fig:f4_control_collapse}, and Appendix
Table~\ref{tab:app_t2_per_seed_deltas}).

\subsection{Cross-dataset consistency of the PARTIAL\_PARTIAL pattern}

The locked criterion-by-criterion outcomes share the same structure
across phases (Appendix Table~\ref{tab:t4_cross_dataset_consistency}):
both close as \texttt{PARTIAL\_EXTERNAL\_REPLICATION}, with
\texttt{GF - Baseline} seed consistency below the 4/5 requirement (1/5,
3/5), \texttt{GF - Gaussian} at 5/5, and the capacity-only and invalidity
triggers inactive. The cross-dataset pattern is therefore
\texttt{PARTIAL\_PARTIAL}, which we report as a consistency of category,
not a cross-dataset upgrade.

\subsection{Supplementary uncertainty and baseline competitiveness}
\label{sec:supp-results}\label{sec:baseline-competitiveness}

These frozen-output analyses do not enter the locked cascade and do not
change the \texttt{PARTIAL\_EXTERNAL\_REPLICATION} outcome (interpretive
boundary in Methods \S\ref{sec:supp}). The A1 paired cluster-bootstrap
95\% CIs on the phase-mean \texttt{GF - Baseline} delta are
\texttt{{[}-0.0064, +0.0160{]}} in Phase C (includes zero; observed
\texttt{+0.0051}) and \texttt{{[}+0.0172, +0.0435{]}} in Phase D (above
zero; observed \texttt{+0.0292}), while the \texttt{GF - Gaussian} and
\texttt{GF - Shuffle} intervals lie far above zero in both phases (all
intervals in Appendix Figure~\ref{fig:f5_bootstrap_ci}). The bootstrap
intervals and the seed-consistency counts answer different questions: in
Phase D the interval lies above zero yet STRONG still fails because only
3/5 seeds clear the +0.020 margin, and in Phase C the interval includes
zero, consistent with the 1/5 count. The A4 audit confirms the
expression baseline is a high-performing in-framework comparator: across
all four \texttt{Baseline - \{MAG-only, Gaussian, Shuffle\}} comparisons
in both phases every locked seed delta is positive (5/5; per-comparison
means in the Appendix). Against this reference the
\texttt{Geneformer - Baseline} gain is modest and seed-variable
(\texttt{+0.0051}, 1/5 in Phase C; \texttt{+0.0292}, 3/5 in Phase D), and
the locked outcome remains \texttt{PARTIAL\_EXTERNAL\_REPLICATION} in both
phases.

\subsection{Cross-architecture consistency and criterion-level spread}
\label{sec:cross-arch}

We now place the Geneformer arm alongside scGPT and UCE under the
same locked nominal downstream evaluation specification (label, controls,
splits, seeds, and cascade; the expression baseline is fitted separately
within each arm's chain); the per-model embedding extraction and its upstream
preprocessing are model-specific and not harmonized
(\S\ref{sec:features}, feature representations). At the outcome-category level the three arms are
consistent: each closes as \texttt{PARTIAL\_EXTERNAL\_REPLICATION} in both
phases (\texttt{PARTIAL\_PARTIAL}), none meets STRONG, and all clear the
matched-capacity Gaussian control (\texttt{Model - Gaussian} mean
$\geq +0.050$ with 5/5 seeds in every arm and phase;
Table~\ref{tab:signed_gb}). This is a consistency result across
architecturally distinct models, not evidence that the protocol
discriminates between them.

\input{tables/main_t5_signed_gb}

The arms separate at the criterion level, at the signed
\texttt{Model - Baseline} delta ($\Delta_{gb}$) that the STRONG rule gates
on (Table~\ref{tab:signed_gb}). In Phase C the Geneformer estimate is
marginally above baseline (\texttt{+0.0051}) but its bootstrap interval
includes zero, so it is not inferentially distinguishable there, while
scGPT and UCE are below baseline (\texttt{-0.0186}, \texttt{-0.0049}). In
Phase D, Geneformer is above baseline with an interval excluding zero
(\texttt{+0.0292}), whereas scGPT and UCE are below (\texttt{-0.0220},
\texttt{-0.0788}). Among the arms with available frozen-output bootstrap
intervals (Geneformer only), Phase D Geneformer is the sole comparison
whose interval lies entirely above zero; the scGPT and UCE point
estimates are negative, but comparable intervals are unavailable, so no
inferential statement is made for those arms. An arm can thus sit clearly
below the baseline (UCE in Phase D, \texttt{-0.0788}, 0/5 seeds above
+0.020) yet still close as \texttt{PARTIAL} by clearing the Gaussian
control (\texttt{+0.1492}, 5/5) without shuffle invalidation:
\texttt{PARTIAL} means above the random-capacity and row-identity
controls, not above the high-performing expression baseline (Appendix
Figure~\ref{fig:signed_gb_forest}).

A scGPT sensitivity arm (a reported-only robustness input that closes as
\texttt{INVALID} in Phase D while leaving the scGPT primary arm
\texttt{PARTIAL}) and the reason bootstrap intervals exist only for the
Geneformer arm are detailed in the Appendix (\emph{Additional
cross-architecture notes}).

%% file: tables/main_t5_signed_gb.tex
\begin{table*}[t]
\centering
\caption{Signed baseline-relative deltas across the three foundation-model
arms under the locked protocol. All three arms close as
\texttt{PARTIAL\_EXTERNAL\_REPLICATION} in both phases (cross-phase
\texttt{PARTIAL\_PARTIAL}); the category label is shared, but the
baseline-relative direction and magnitude differ across arms. $\Delta_{gb}$
is the mean test-$R^2$ difference between the frozen-representation arm and
the strong expression baseline; ``seeds'' counts locked training seeds with
per-seed $\Delta_{gb}\geq+0.020$ (STRONG requires $\geq 4/5$). Bootstrap CIs
are the frozen-output A1 cluster-bootstrap 95\% intervals; they are available
only for the Geneformer arm (see note).}
\label{tab:signed_gb}
\small
\begin{tabular}{llccccc}
\toprule
Phase & Arm & mean($R^2$) & Baseline$^{\dagger}$ & $\Delta_{gb}$ & seeds$\geq$+0.020 & $\Delta_{gb}$ 95\% CI \\
\midrule
C (RPE1) & Geneformer & 0.8362 & 0.8310 & $+0.0051$ & 1/5 & $[-0.0064,\,+0.0160]$ \\
C (RPE1) & scGPT      & 0.8156 & 0.8343 & $-0.0186$ & 1/5 & n/a$^{\ddagger}$ \\
C (RPE1) & UCE        & 0.8294 & 0.8343 & $-0.0049$ & 1/5 & n/a$^{\ddagger}$ \\
\midrule
D (K562) & Geneformer & 0.7944 & 0.7652 & $+0.0292$ & 3/5 & $[+0.0172,\,+0.0435]$ \\
D (K562) & scGPT      & 0.7387 & 0.7607 & $-0.0220$ & 0/5 & n/a$^{\ddagger}$ \\
D (K562) & UCE        & 0.6819 & 0.7607 & $-0.0788$ & 0/5 & n/a$^{\ddagger}$ \\
\bottomrule
\end{tabular}
\vspace{2pt}
\caption*{\footnotesize
$^{\dagger}$ $\Delta_{gb}$ is computed against each arm's own committed
Baseline branch; the Geneformer Baseline differs from the common
scGPT/UCE Baseline by $\approx 0.003$ (C) and $\approx 0.005$ (D), which
changes no locked outcome (no cross-arm renormalization is applied).
$^{\ddagger}$ Row-level predictions were not persisted for the scGPT and
UCE final-test runs, so frozen-output bootstrap CIs are unavailable; those
two arms are reported descriptively, not with interval-based uncertainty.}
\end{table*}

%% file: sections/04_discussion.tex
\section{Discussion}

Under this pre-registered, final-test-once framework, three
architecturally distinct frozen representations (Geneformer, scGPT, and
UCE) close as \texttt{PARTIAL\_EXTERNAL\_REPLICATION} in both external
Replogle Perturb-seq phases (\texttt{PARTIAL\_PARTIAL}), locating a
transfer boundary rather than a binary outcome: the representations carry
signal well beyond the matched-capacity and row-identity controls
(Results \S\ref{sec:control}), yet the gain over a strong expression
baseline is not consistent enough across seeds to meet the
pre-registered STRONG criterion (\texttt{GF - Baseline} clears +0.020 in
1/5 seeds in Phase C and 3/5 in Phase D, below the 4/5 gate). A PARTIAL outcome under
pre-registration is informative precisely because it identifies where
transfer holds and where it does not, preventing both overclaiming and
unwarranted dismissal; and because the structure recurs in both
datasets, \texttt{PARTIAL\_PARTIAL} is a consistency finding, not a
cross-dataset upgrade to STRONG.

Across Geneformer, scGPT, and UCE the outcome category is consistent
(every arm clears the Gaussian control and none meets STRONG;
Table~\ref{tab:signed_gb}), which is a consistency result, not evidence
that the protocol discriminates between architectures. The arms separate
only at the criterion level, where the signed \texttt{Model - Baseline}
delta differs in sign and magnitude (\S\ref{sec:cross-arch}). An arm
below the baseline, as UCE is in Phase D, can still close as
\texttt{PARTIAL} by clearing the Gaussian and Shuffle controls;
\texttt{PARTIAL} therefore means ``above the random-capacity and
row-identity controls,'' not ``beats the high-performing expression
baseline.''

The framework separates three often-conflated questions (non-random
signal, gain over a strong baseline, and seed-stability under a
pre-registered rule), so its contribution stands independently of any
single empirical outcome. It is thus a reproducible, boundary-finding
evaluation framework that produced an honest, bounded outcome, not a
STRONG-replication claim.

%% file: sections/05_limitations.tex
\section{Limitations}
\label{sec:limitations-supp}

The external evidence comprises two locked phases on two related Replogle
Perturb-seq datasets (RPE1 and K562) from the same study family, so
\texttt{PARTIAL\_PARTIAL} characterizes consistency across two related
Perturb-seq settings, not across other tissue contexts, perturbation
modalities, assay platforms, or CRISPR technologies. This is a bounded
case study, not a general statement about foundation models.

The weak label is a per-row response-magnitude proxy, not
direction-specific mechanism or phenotype, and its zero-clamp and
\texttt{log1p} reduce information at the weak-effect end. Because both the
label and the MAG anchor are magnitude-based, the anchor already captures
most achievable in-framework performance: on validation (Geneformer arm,
Appendix Table~\ref{tab:app_t1_per_seed_metrics}), \texttt{MAG-only} reaches
mean $R^2 = 0.717 / 0.783$ (Phase C/D) against a high-performing
comparator (the expression baseline, not a task ceiling) of
$0.854 / 0.881$, leaving only $\approx 0.137 / 0.098$ between them. A
modest gain over the baseline is thus consistent with a largely
magnitude-driven task; whether the modesty reflects the representation or
a ceiling property of the label is unresolved, and stronger labels (for
example perturbation-specific phenotype scores) would refine the boundary.

The locked STRONG gate ($\geq$4/5 seeds with \texttt{GF - Baseline}
$\geq$ +0.020) is deliberately an optimization-robustness criterion,
independent of the bootstrap interval; the two can point in different
directions (in Phase D the interval lies above zero yet STRONG fails at
3/5 seeds). This is one pre-registered choice, and the defense is the
commit that fixed the rule before any test-set inspection. The locked
five-model family and its fixed hyperparameters, epochs, and seeds are
pre-registered constraints, not tuning opportunities; the matched-capacity
Gaussian and within-split Shuffle controls provide evidence against the
specified random-capacity and row-identity explanations, but no finite
control set excludes every confound. The supplementary analyses A1--A4
operate on committed frozen predictions and do not enter the locked
cascade.

Two further caveats deserve emphasis. First, \emph{external} here
denotes externality relative to the downstream training and split
construction, not to foundation-model pretraining, whose overlap with
the Replogle datasets cannot be excluded without full pretraining
manifests. Second, each arm uses separately committed \texttt{Baseline}
and \texttt{MAG-only} chains, so cross-arm comparisons are descriptive and
we do not rank architectures (six further limitations: Appendix,
\emph{Extended limitations}).

%% file: appendix/appendix.tex
\section*{Appendix}
\renewcommand{\thetable}{A\arabic{table}}
\renewcommand{\thefigure}{A\arabic{figure}}
\setcounter{table}{0}
\setcounter{figure}{0}

\subsection*{Frozen-output paired bootstrap CI forest plot}

\begin{figure}[ht]
\centering
\includegraphics[width=0.85\textwidth]{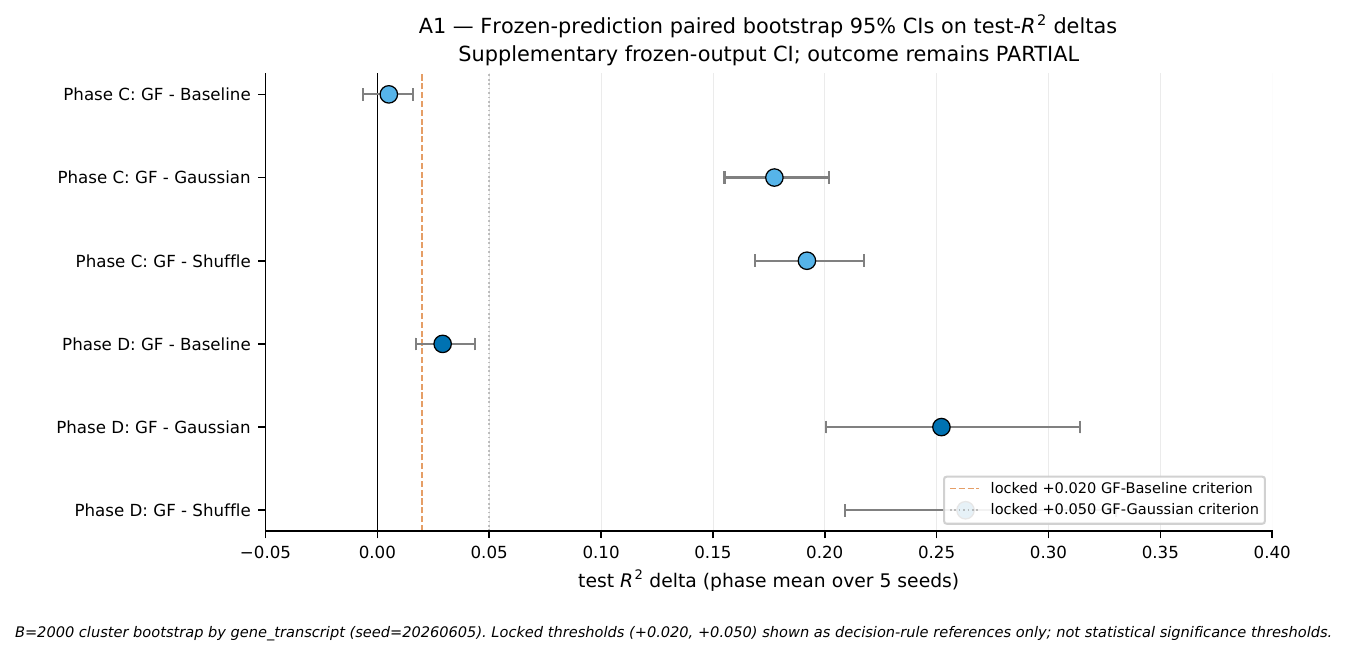}
\caption{Paired cluster-bootstrap 95\% confidence intervals on test-$R^2$ deltas (frozen-output supplementary analysis A1). $B = 2{,}000$ resamples; bootstrap seed \texttt{20260605}; bootstrap unit = unique \texttt{gene\_transcript}. The intervals quantify row-level test uncertainty under cluster resampling; they do not replace the locked decision-rule cascade. See the interpretive boundary in Methods \S\ref{sec:supp} and Results \S\ref{sec:supp-results}. These intervals are available for the Geneformer arm because its final-test run persisted seed-level row predictions; the scGPT and UCE final-test runners persisted only summary metrics, so those arms are reported without frozen-output confidence intervals (Table~\ref{tab:signed_gb}).}
\label{fig:f5_bootstrap_ci}
\end{figure}

\subsection*{Artifact fingerprints and manifest details}

This subsection consolidates the cryptographic fingerprints, random
seeds, and label-manifest internals referenced throughout the main
text. All values are recorded in the committed final-test QC artifacts
and were not modified after the locked final-test evaluation.

\textbf{SHA-256 fingerprints (Phase D primary).}
The Phase D primary D1 label parquet has SHA-256
\texttt{\seqsplit{79d5453bc1854c32019469e9cf9642920615effce85340d3c1db82d60170b342}}.
The Phase D split assignment has SHA-256
\texttt{\seqsplit{31fd1c6b032557c82d88f7664c34cd0627a89eff598ff9ad99ff17b2ce2f39e1}}.
Both values were re-verified immediately before and after the locked
final-test evaluation.

\textbf{Random seeds.}
The five locked model seeds are
\texttt{{[}20260527, 20260528, 20260529, 20260530, 20260531{]}}; the
Phase C split seed is \texttt{20260601}; the Phase D split seed is
\texttt{20260602}; and the supplementary-bootstrap seed is
\texttt{20260605}.

\textbf{Label-manifest internals.}
The Phase D D1 label-construction manifest records
\texttt{raw\_counts\_used = true}, \texttt{normalized\_counts\_used = false},
and \texttt{label\_formula\_version = phase\_c\_consistent\_global\_nt\_ad\_v1},
and attests that the Phase C label-generation logic was mirrored
exactly (same \texttt{scipy} 1.15.2, same \texttt{anderson\_ksamp} $k = 2$,
finite-only mean, and \texttt{log1p(max(0, y))} clip). The frozen
Geneformer-V2-104M (gc104M vocabulary) representation uses a
768-dimensional embedding in both phases, and the matched-capacity
Gaussian control uses the same 768-dimensional capacity in both phases.

\subsection*{Additional Geneformer-arm figures and tables (moved from the main text for space)}

\begin{figure}[ht]
\centering
\includegraphics[width=0.9\textwidth]{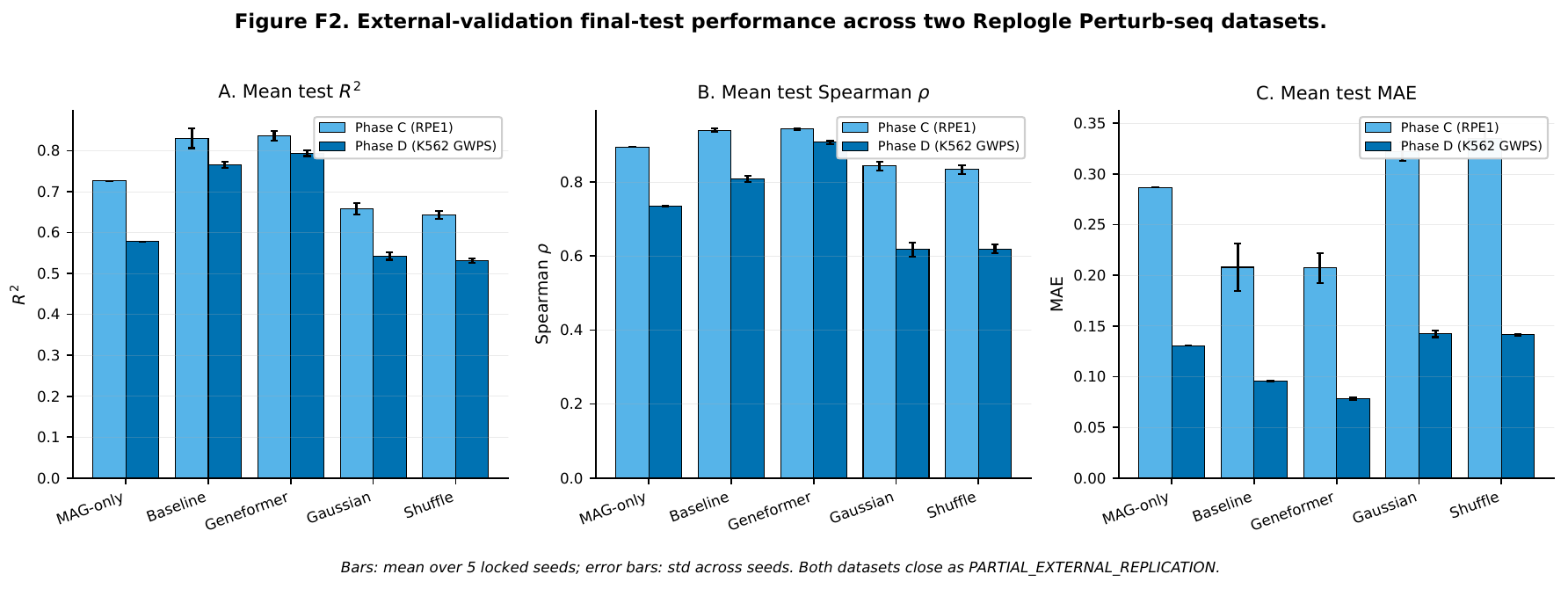}
\caption{Final-test performance for the \textbf{Geneformer arm's} five-model family in both external phases (Phase C: RPE1; Phase D: K562 GWPS), shown in full as the primary worked example. Bars show mean test $R^2$ across the five locked training seeds for each of \texttt{MAG-only}, \texttt{Baseline}, \texttt{Geneformer}, \texttt{Gaussian}, and \texttt{Shuffle}; error bars show seed-level standard deviation. The controls collapse well below Baseline in both phases, confirming control validity. The three-arm baseline-relative comparison is given in Figure~\ref{fig:signed_gb_forest} and Table~\ref{tab:signed_gb}.}
\label{fig:f2_test_performance}
\end{figure}

\begin{figure}[ht]
\centering
\includegraphics[width=0.9\textwidth]{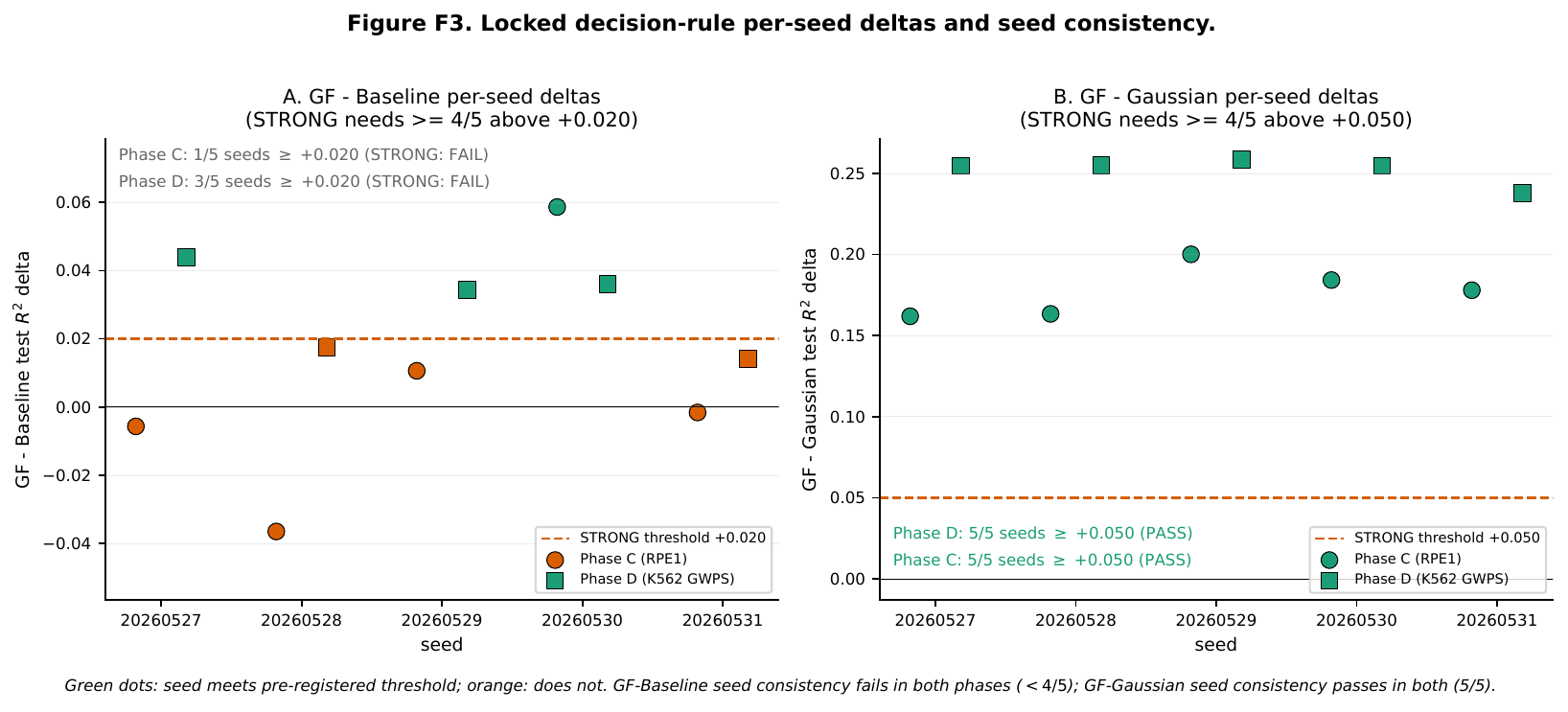}
\caption{Locked decision-rule per-seed deltas for the \textbf{Geneformer arm}, shown in full as the primary worked example. Per-seed $R^2$ deltas for GF$-$Baseline and GF$-$Gaussian are plotted against the locked decision thresholds ($+0.020$ and $+0.050$), with seed-consistency counts (the 4/5 STRONG criterion) shown for each comparator. Per-seed deltas at this resolution are available for the Geneformer arm because its final-test run persisted seed-level predictions; for the scGPT and UCE arms the final-test runners persisted only summary metrics, so their arms are summarized by signed baseline-relative deltas and seed-gate counts in Table~\ref{tab:signed_gb} and Figure~\ref{fig:signed_gb_forest} rather than at per-seed resolution.}
\label{fig:f3_decision_rule_deltas}
\end{figure}

\begin{figure}[ht]
\centering
\includegraphics[width=0.9\textwidth]{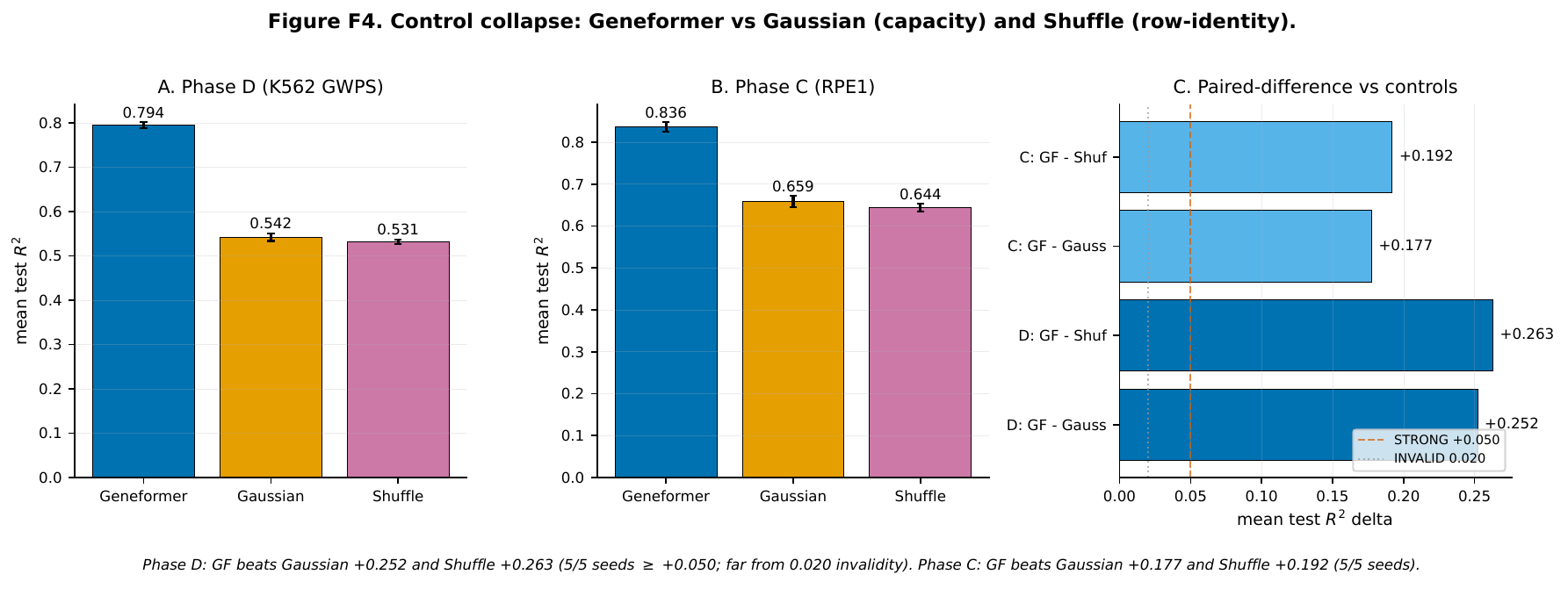}
\caption{Control collapse for the \textbf{Geneformer arm}, shown in full as the primary worked example. Matched-capacity Gaussian and within-split Shuffle controls collapse to substantially lower test $R^2$ than both Geneformer and the expression baseline in both external phases. Paired GF$-$Gaussian and GF$-$Shuffle deltas are reported alongside the Gaussian$-$Baseline difference (negative in both phases, indicating that random-capacity features alone do not beat the strong expression baseline). The same control collapse holds for the scGPT and UCE arms; their full five-model families are reported in Appendix Table~\ref{tab:app_t4_scgpt_uce_family}, and the three-arm baseline-relative comparison in Table~\ref{tab:signed_gb} and Figure~\ref{fig:signed_gb_forest}.}
\label{fig:f4_control_collapse}
\end{figure}

\input{tables/main_t1_dataset_split}

\input{tables/main_t2_final_performance}

\input{tables/main_t3_decision_rule}

\input{tables/main_t4_cross_dataset}

\begin{figure}[ht]
\centering
\includegraphics[width=0.9\textwidth]{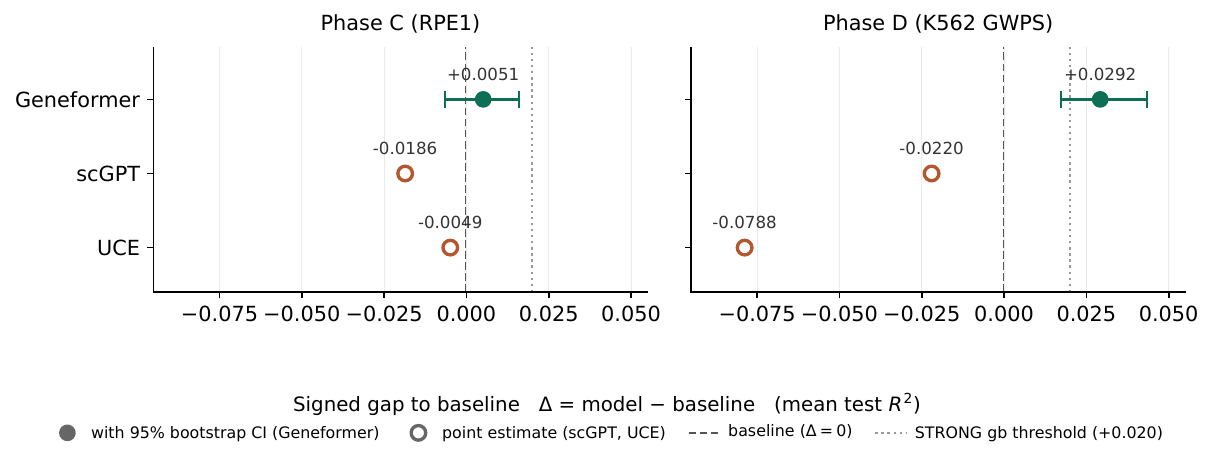}
\caption{Signed baseline-relative gap for all three frozen representations,
as a visual companion to Table~\ref{tab:signed_gb}. Each point is
$\Delta = \text{model} - \text{baseline}$ in mean test $R^2$ (five locked
seeds) against that arm's own baseline; the dashed line marks the baseline
($\Delta=0$) and the dotted line the pre-registered STRONG gb threshold
($+0.020$). Geneformer is shown with a bootstrap 95\% CI because its
committed final-test artifacts preserve row-level predictions; scGPT and UCE
are shown as point estimates because their committed artifacts preserve
summary metrics and seed-gate counts but not row-level predictions. Figure~\ref{fig:signed_gb_forest}
is therefore a descriptive cross-arm comparison, not an interval-based
ranking figure. All three arms sit near the baseline and none reaches the STRONG threshold;
only the Geneformer Phase D interval clears and excludes zero, while in
Phase C its interval includes zero. This illustrates that
\texttt{PARTIAL\_EXTERNAL\_REPLICATION} means above random controls but not
reliably above the strong expression baseline.}
\label{fig:signed_gb_forest}
\end{figure}

\subsection*{Additional cross-architecture notes}

\textbf{A4 baseline-competitiveness means.} The mean
\texttt{Baseline - \{MAG-only, Gaussian, Shuffle\}} test-$R^2$ gains are
\texttt{+0.1041 / +0.1724 / +0.1869} in Phase C and
\texttt{+0.1874 / +0.2230 / +0.2337} in Phase D; every locked seed delta
is positive (5/5 per comparison).

For pre-specified transparency we report the scGPT sensitivity arm, a
reported-only robustness input and not the scGPT primary outcome. Under
that input form Phase D closes as \texttt{INVALID} because the
within-split shuffle of the scGPT delta sits at the anchor level
(\texttt{abs(Shuffle - comparator) < 0.020}), so row-identity information
does not separate signal from artifact for that input form; the scGPT
primary arm is unaffected (\texttt{abs(Shuffle - comparator) = 0.1637})
and remains \texttt{PARTIAL\_EXTERNAL\_REPLICATION}, with no role swap
between primary and sensitivity.

Frozen-output bootstrap intervals are available only for the Geneformer
arm, because row-level test predictions were persisted only for that arm
at final-test time; the scGPT and UCE final-test runners persisted
per-seed summary metrics but not row-level predictions, so intervals
cannot be computed for those arms from the committed artifacts, and the
locked test partition was deliberately not re-touched to generate them.
For scGPT and UCE we therefore report per-seed deltas and seed-gate
counts, which are available for all arms, and we avoid interval-based
cross-arm inference for those two arms.

\subsection*{Appendix Tables}

\input{appendix/appendix_t1_per_seed_metrics}
\input{appendix/appendix_t2_per_seed_deltas}
\input{appendix/appendix_t3_artifact_lineage}
\input{tables/app_t4_scgpt_uce_family}

\input{appendix/appendix_a4_allowed_disallowed_claims}
\input{appendix/appendix_t5_repro_checklist}

\subsection*{Supplementary methods-note pointer}

The supplementary robustness-analysis documentation accompanying this
manuscript describes the cluster-bootstrap implementation used by the
supplementary A1 analysis, including the bootstrap unit, the number of
replicates ($B = 2{,}000$), the bootstrap seed (\texttt{20260605}), the
strict per-replicate pairing assertion on the test prediction arrays,
the phase-level summary defined by averaging per-seed bootstrap deltas
at each replicate and taking the $[2.5, 97.5]$ percentile interval, and
the cluster-size-1 caveat that cluster resampling reduces to paired
row-level resampling in these evaluated test sets. The same interpretive-boundary wording appears in Methods
\S\ref{sec:supp} and Results \S\ref{sec:supp-results} of the main paper.

\subsection*{Extended methods details (moved from the main text)}

\textbf{Anchor selection grid.} The locked \texttt{RidgeCV}-style MAG
anchor sweeps \texttt{ALPHA\_GRID = {[}1e-4, 1e-3, 1e-2, 1e-1, 1.0, 10.0,
100.0, 1000.0{]}} under train-only \texttt{GroupKFold(5)} grouped by
\texttt{target\_gene}; both phases independently selected the anchor
penalty \texttt{$\alpha$ = 0.01}, identically across arms. This anchor
penalty is distinct from the expression \texttt{Baseline} branch, a
separate \texttt{Ridge} fit whose Phase~C penalty differs across arms
(\emph{Separate baseline execution chains} under Extended limitations);
the two selections should not be conflated.

\textbf{Per-split unique \texttt{target\_gene} counts.} Phase D:
6{,}757 / 1{,}448 / 1{,}448 unique \texttt{target\_gene} values for
train / validation / test; Phase C: 1{,}508 / 323 / 323.

\textbf{A1 bootstrap statistic.} For each
\texttt{(phase $p$, comparator $c \in$ \{Baseline, Gaussian, Shuffle\},
seed $s$, replicate $b$)}, unique \texttt{gene\_transcript} values are
resampled with replacement and a paired test-$R^2$ delta is computed on
the resampled rows as

\begin{equation*}
\delta_{p,c,s,b} \;=\; R^2\!\left(y_{\text{true},b},\ \hat{y}^{\,\text{GF}}_{b}\right) \;-\; R^2\!\left(y_{\text{true},b},\ \hat{y}^{\,c}_{b}\right),
\end{equation*}

with a local vectorized $R^2$ numerically equivalent to
\texttt{sklearn.metrics.r2\_score}. Per-seed 95\% confidence intervals are
the \texttt{{[}2.5, 97.5{]}} percentiles of
$\{\delta_{p,c,s,b}\}_{b=1}^{2000}$; the phase-mean interval is obtained
by averaging per-seed bootstrap deltas within each replicate and taking
the same percentiles.

\subsection*{scGPT and UCE extraction specifications}
\label{sec:fm-extraction}

The Geneformer extraction is specified in the main text
(\S\ref{sec:features} feature-representation paragraph). The scGPT and UCE
per-cell embedding-extraction configurations are read verbatim from the
committed extraction scripts (Appendix
Table~\ref{tab:app_t3_artifact_lineage}) and are reported below. These
upstream steps are model-specific and are not harmonized across arms; we
report them so that the ``downstream evaluation specification is held
fixed'' claim can be checked against the exact per-model preprocessing.

\footnotesize
\renewcommand{\arraystretch}{1.2}
\begin{tabularx}{\textwidth}{@{}>{\raggedright\arraybackslash}p{2.9cm}>{\raggedright\arraybackslash}X>{\raggedright\arraybackslash}X@{}}
\toprule
Item & scGPT & UCE \\
\midrule
Checkpoint & \texttt{whole\_human/best\_model.pt} (CELLxGENE census human, snapshot \texttt{cellxgene\_census\_human-May23-08-36-2023}) & \texttt{33l\_8ep\_1024t\_1280.torch} (33-layer; figshare article 24320806, v5) \\
Checkpoint SHA-256 & \texttt{\seqsplit{6cb5d451ab5c4b33eb673adbe4fddc61d2389df1b89b7651a9fe2e557572b922}} & \texttt{\seqsplit{3f458726196308e171611ed28394b55865708749f82af68cfd7771d5dfee661e}} (MD5 matches figshare) \\
Public source / license & \url{https://github.com/bowang-lab/scGPT} (code MIT) & \url{https://github.com/snap-stanford/UCE} (code MIT); weights figshare (CC BY 4.0) \\
Embedding dimension & 512 & 1280 \\
Gene-matching policy & Ensembl$\rightarrow$HUGO via inverted \texttt{gene\_name\_id\_dict\_gc104M} (same bridge as UCE); matched to the 60{,}697-token scGPT HUGO vocabulary & Upstream requires gene symbols (not Ensembl); same Ensembl$\rightarrow$symbol bridge \\
Input genes / coverage & Phase C 8{,}628/8{,}749 (98.62\%); Phase D 8{,}076/8{,}248 (97.91\%); mean per-cell OOV 0.78\% (C), 1.20\% (D) & Phase C 8{,}741/8{,}749 and Phase D 8{,}243/8{,}248 retained after Ensembl$\rightarrow$symbol bridging; exact post-bridge UCE/ESM2 token coverage was not persisted \\
OOV handling & Drop unmapped Ensembl and OOV HUGO before top-$K$ & Drop unmapped Ensembl at bridge; post-bridge OOV handled by the upstream UCE evaluator \\
Normalization / binning & Raw counts (no library-size or log transform); scGPT binning with \texttt{n\_bins=51}, per-cell seed \texttt{(20260527 + orig\_obs\_idx) mod (2$^{32}-1$)}, NumPy RNG state saved/restored per call & Raw counts; no library-size normalization or binning. \texttt{log1p(raw count)} defines weighted gene-sampling probabilities only; expression values are not passed to the model \\
Batch / domain / species & None (\texttt{use\_batch\_labels=False}) & \texttt{--species human} \\
Extraction layer / tensor & Encoder output \texttt{cell\_emb} (\texttt{model.\_encode}) & Output of the frozen 33-layer Transformer encoder and decoder \\
Pooling & CLS position \texttt{cell\_emb[:,0,:]}, padding masked; 512-D cell embedding L2-normalized & CLS-position decoder output \texttt{gene\_output[0,:,:]}, padding excluded by the key-padding mask; 1280-D cell embedding L2-normalized \\
Sequence / truncation & 1200: deterministic top-1199 genes by raw-count descending (tie-break ascending vocab index) plus CLS & 1{,}024 genes sampled with replacement (probability $\propto \log(1{+}x)$); chromosome boundary tokens added; padded to \texttt{pad\_length=1536}; no deterministic top-$K$ \\
Gene ordering & After deterministic top-$K$; CLS prepended & Chromosome-organized: chromosome order shuffled under the locked seed, genes within a chromosome sorted by genomic start \\
Precision & \texttt{autocast} enabled; output cast to FP32; L2 row-normalized & FP32 output; no explicit mixed-precision override recorded \\
Per-row aggregation & \multicolumn{2}{>{\raggedright\arraybackslash}X}{\texttt{mean(emb)\_pert $-$ mean(emb)\_NT} (phase-global NT mean), then L2-unit-normalized; identical form across Geneformer, scGPT, and UCE} \\
Extraction-script commit & \texttt{213376e} (P2-B1) & \texttt{7ab7bc8} (P2-D3.5B) \\
\bottomrule
\end{tabularx}

\noindent The UCE rows are recovered from the committed upstream call
graph (\texttt{snap-stanford/UCE}, \texttt{eval\_single\_anndata.py} /
\texttt{evaluate.py} / \texttt{eval\_data.py}); the only value not
recovered from committed local evidence is the exact post-bridge
UCE/ESM2 token coverage. The two models construct model input
differently: scGPT discretizes expression into 51 bins, whereas UCE
passes no expression values to the model and instead samples 1{,}024
genes with probability proportional to \texttt{log1p} of the raw count.
Cross-arm differences therefore reflect the encoder together with its
model-specific preprocessing, not the encoder alone.

\subsection*{Extended limitations}
\label{sec:extended-limitations}

The following extend the limitations stated in the main text.

\begin{enumerate}
\setlength{\itemsep}{1pt}
\item \textbf{Pretraining-overlap is unaudited.} We use \emph{external}
in the sense of externality relative to the downstream supervised
training and split construction. We cannot exclude overlap between the
foundation-model pretraining corpora and the Replogle datasets: the
committed scGPT \texttt{whole\_human} configuration records a CELLxGENE
census human corpus (snapshot \texttt{cellxgene\_census\_human-May23-08-36-2023}),
which post-dates Replogle 2022 and could ingest it, and the UCE
(snap-stanford) pretraining corpus is not documented locally at the
study level; neither corpus explicitly lists or excludes Replogle. A
per-model audit against complete pretraining manifests is left for future
work.
\item \textbf{Same study family.} Both phases are Replogle 2022
genome-wide CRISPRi Perturb-seq datasets; externality across laboratory,
assay platform, and study family is not evaluated.
\item \textbf{Separate baseline execution chains.} The intended cross-arm
comparability is specification-level, not byte-level. The \texttt{Baseline}
branch is re-fit within each arm's committed chain rather than reused as
a single frozen prediction table. The chains share the same committed
label, split assignment, seeds, \texttt{ALPHA\_GRID}, and training
specification, and the \texttt{MAG-only} anchor is byte-identical across
arms (Phase D \texttt{MAG-only} test-$R^2 = 0.5778$), yet the executed
baselines differ: the Phase C chains selected different train-only Ridge
penalties (\texttt{$\alpha$} = 0.01 for Geneformer, 0.001 for scGPT/UCE)
despite nominally identical grids and grouping, and the Phase D Baseline
mean test-$R^2$ differs (0.7652 vs 0.7607). The committed records do not
isolate the numerical cause of these differences. Each arm's
baseline-relative delta is therefore computed against its own committed
\texttt{Baseline}; cross-arm comparisons are descriptive and we do not
rank architectures on absolute test-$R^2$, and none of these differences
changes any locked outcome category.
\item \textbf{Bootstrap unit.} The frozen-output bootstrap resamples at
the \texttt{gene\_transcript} (row) level; because the split is grouped
at \texttt{target\_gene}, a \texttt{target\_gene}-clustered bootstrap
would be more conservative. Row predictions are unavailable for scGPT and
UCE, so no bootstrap is reported for those arms.
\item \textbf{Coarse seed gate.} With five seeds the $\geq$4/5
seed-consistency gate is discontinuous: a single seed changes the pass
rate by 20 percentage points.
\item \textbf{AD-label sample-size sensitivity.} The Anderson-Darling
statistic can depend on the number of perturbed cells per row; we did not
audit the association between per-row cell count and label magnitude, nor
fully resolve possible gem-group or batch confounding in the global
non-targeting pool.
\item \textbf{Frozen embeddings only.} All representations are used
frozen, with no fine-tuning; checkpoint selection may affect conclusions.
\item \textbf{Model-specific preprocessing not harmonized.} As noted in
the feature-representation methods, per-model tokenization, vocabulary,
gene mapping, normalization, pooling, and sequence handling differ across
arms (see \emph{scGPT and UCE extraction specifications} above).
\end{enumerate}

%% file: tables/main_t1_dataset_split.tex
\begin{table}[t]
\caption{Dataset and split design summary for the two external Replogle Perturb-seq phases (Phase C: RPE1; Phase D: K562 GWPS). Full path-resolved provenance is reported in Appendix Table~\ref{tab:app_t3_artifact_lineage}.}
\label{tab:t1_dataset_split}
\centering
\footnotesize
\setlength{\tabcolsep}{4pt}
\renewcommand{\arraystretch}{1.15}
\begin{tabularx}{\textwidth}{@{}llrrrr>{\raggedright\arraybackslash}X>{\raggedright\arraybackslash}X@{}}
\toprule
Phase & Cell line & $n_{\mathrm{rows}}$ & $n_{\mathrm{train}}$ & $n_{\mathrm{val}}$ & $n_{\mathrm{test}}$ & Split policy & Final outcome \\
\midrule
Phase C & RPE1 & 2{,}301  & 1{,}612 & 344   & 345   & \texttt{target\_gene}-grouped, 70/15/15 (seed 20260601) & \texttt{PARTIAL\_\allowbreak EXTERNAL\_\allowbreak REPLICATION} \\
\addlinespace[2pt]
Phase D & K562 & 10{,}439 & 7{,}320 & 1{,}554 & 1{,}565 & \texttt{target\_gene}-grouped, 70/15/15, no stratification (seed 20260602) & \texttt{PARTIAL\_\allowbreak EXTERNAL\_\allowbreak REPLICATION} \\
\bottomrule
\end{tabularx}
\end{table}

%% file: tables/main_t2_final_performance.tex
\begin{table}[t]
\caption{Final-test performance (mean $\pm$ std across the five locked training seeds) for the \textbf{Geneformer arm's} five-model family in Phase C and Phase D; this table presents the Geneformer arm in full as the primary worked example. Path-resolved source artifacts are reported in Appendix Table~\ref{tab:app_t3_artifact_lineage}. The corresponding full model families for the scGPT and UCE arms are reported in Appendix Table~\ref{tab:app_t4_scgpt_uce_family}, and the three-arm baseline-relative comparison is given in Table~\ref{tab:signed_gb}.}
\label{tab:t2_final_performance}
\centering
\scriptsize
\begin{tabular}{@{}llrrrrrr@{}}
\toprule
Phase & Model & $\overline{R^2}$ & $\sigma_{R^2}$ & $\overline{\rho}$ & $\sigma_{\rho}$ & $\overline{\mathrm{MAE}}$ & $\sigma_{\mathrm{MAE}}$ \\
\midrule
Phase C & C-MAG-only & 0.7269 & 0.0000 & 0.8953 & 0.0000 & 0.2865 & 0.0000 \\
Phase C & C-Baseline & 0.8310 & 0.0244 & 0.9400 & 0.0048 & 0.2080 & 0.0237 \\
Phase C & C-Geneformer & 0.8362 & 0.0115 & 0.9436 & 0.0024 & 0.2071 & 0.0146 \\
Phase C & C-Gaussian & 0.6587 & 0.0136 & 0.8437 & 0.0116 & 0.3221 & 0.0090 \\
Phase C & C-Shuffle & 0.6441 & 0.0096 & 0.8344 & 0.0120 & 0.3354 & 0.0099 \\
Phase D & D-MAG-only & 0.5778 & 0.0000 & 0.7351 & 0.0000 & 0.1305 & 0.0000 \\
Phase D & D-Baseline & 0.7652 & 0.0075 & 0.8090 & 0.0088 & 0.0957 & 0.0006 \\
Phase D & D-Geneformer & 0.7944 & 0.0071 & 0.9078 & 0.0040 & 0.0782 & 0.0015 \\
Phase D & D-Gaussian & 0.5422 & 0.0087 & 0.6182 & 0.0187 & 0.1422 & 0.0032 \\
Phase D & D-Shuffle & 0.5315 & 0.0051 & 0.6195 & 0.0123 & 0.1414 & 0.0007 \\
\bottomrule
\end{tabular}
\end{table}

%% file: tables/main_t3_decision_rule.tex
\begin{table}[t]
\caption{Locked decision-rule evaluation: per-criterion thresholds, observed values, and pass/fail status for both phases. The \emph{GF\,$-$\,Baseline seed-consistency} row is the gating criterion that keeps STRONG locked out in both phases. The capacity-only and invalidity triggers are inactive in both phases. \textbf{Priority cascade:} INVALID $>$ STRONG $>$ CAPACITY\_ONLY $>$ PARTIAL $>$ NO. Final outcome assigned in both phases: \texttt{PARTIAL\_EXTERNAL\_REPLICATION}; cross-dataset pattern: \texttt{PARTIAL\_PARTIAL}. This cascade is shown in full for the Geneformer arm as the primary worked example; the corresponding signed baseline-relative deltas and seed-gate counts for the scGPT and UCE arms are given in Table~\ref{tab:signed_gb} and Figure~\ref{fig:signed_gb_forest}.}
\label{tab:t3_decision_rule}
\centering
\footnotesize
\setlength{\tabcolsep}{5pt}
\renewcommand{\arraystretch}{1.25}
\begin{tabular}{@{}>{\raggedright\arraybackslash}p{4.7cm}>{\raggedright\arraybackslash}p{3.4cm}rcrc@{}}
\toprule
\textbf{Criterion} & \textbf{Locked threshold} & \multicolumn{2}{c}{\textbf{Phase C}} & \multicolumn{2}{c}{\textbf{Phase D}} \\
\cmidrule(lr){3-4} \cmidrule(lr){5-6}
 &  & Observed & P/F & Observed & P/F \\
\midrule
GF $-$ Baseline mean $R^2$              & $\geq$ +0.020                                      & +0.00513   & \textbf{FAIL} & +0.02918   & \textbf{PASS} \\
GF $-$ Baseline seed consistency        & $\geq$ 4/5 seeds with $\Delta \geq$ +0.020         & 1/5        & \textbf{FAIL} & 3/5        & \textbf{FAIL} \\
GF $-$ Gaussian mean $R^2$              & $\geq$ +0.050                                      & +0.17748   & \textbf{PASS} & +0.25213   & \textbf{PASS} \\
GF $-$ Gaussian seed consistency        & $\geq$ 4/5 seeds with $\Delta \geq$ +0.050         & 5/5        & \textbf{PASS} & 5/5        & \textbf{PASS} \\
Gaussian $-$ Baseline (capacity check)  & mean $\Delta \geq$ +0.010                          & $-$0.17236 & \textit{(not triggered)} & $-$0.22295 & \textit{(not triggered)} \\
$|$Shuffle $-$ Geneformer$|$ (invalidity check) & $<$ 0.020                                  & 0.19201    & \textit{(not triggered)} & 0.26290    & \textit{(not triggered)} \\
\bottomrule
\end{tabular}
\end{table}

%% file: tables/main_t4_cross_dataset.tex
\begin{table}[t]
\caption{Cross-dataset consistency matrix for the \textbf{Geneformer arm}, supporting the \texttt{PARTIAL\_PARTIAL} pattern (source: supplementary A3), shown in full as the primary worked example. Each row reports a criterion's status side-by-side for Phase C and Phase D. Every row's interpretation guardrail is identical and stated once: \emph{Consistent PARTIAL pattern; not an upgrade to STRONG.} \texttt{P\_E\_R} = \texttt{PARTIAL\_EXTERNAL\_REPLICATION}. The scGPT and UCE arms independently reproduce the same \texttt{PARTIAL\_PARTIAL} outcome across both phases; their signed baseline-relative deltas and seed-gate counts are given in Table~\ref{tab:signed_gb} and Figure~\ref{fig:signed_gb_forest}.}
\label{tab:t4_cross_dataset_consistency}
\centering
\footnotesize
\setlength{\tabcolsep}{4pt}
\renewcommand{\arraystretch}{1.18}
\begin{tabularx}{\textwidth}{@{}>{\raggedright\arraybackslash}p{4.0cm}>{\raggedright\arraybackslash}X>{\raggedright\arraybackslash}X>{\raggedright\arraybackslash}X@{}}
\toprule
\textbf{Evidence dimension} & \textbf{Phase C value} & \textbf{Phase D value} & \textbf{Consistency pattern} \\
\midrule
Assigned outcome & \texttt{P\_E\_R} & \texttt{P\_E\_R} & Both \texttt{P\_E\_R} (consistent). \\
Cross-dataset pattern & PARTIAL & PARTIAL & \texttt{PARTIAL\_PARTIAL} (consistent). \\
GF $-$ Baseline mean $R^2$ delta & +0.00513 & +0.02918 & Both $>0$; Phase C below +0.020; Phase D above. \\
GF $-$ Baseline seed consistency ($\geq$ +0.020 / 5) & 1/5 (STRONG needs 4/5) & 3/5 (STRONG needs 4/5) & Both below 4/5; STRONG locked out (consistent). \\
GF $-$ Gaussian mean $R^2$ delta & +0.17748 & +0.25213 & Both far above +0.050; matched-capacity advantage clear. \\
GF $-$ Gaussian seed consistency ($\geq$ +0.050 / 5) & 5/5 & 5/5 & Both 5/5; PASS in both. \\
Capacity-only trigger (Gaussian $-$ Baseline $\geq$ +0.010) & $-$0.17236 (not triggered) & $-$0.22295 (not triggered) & Both negative; capacity-only ruled out. \\
Invalid-shuffle trigger ($|$Shuffle $-$ Geneformer$|$ $<$ 0.020) & 0.19201 (not triggered) & 0.26290 (not triggered) & Both far above 0.020; shuffle does not invalidate. \\
Final-test-once and no rerun & yes (val parity 1.11e$-$16) & yes (val parity 1.11e$-$16; test evaluated once = true) & Final-test-once preserved in both. \\
D1.5 used for outcome & N/A (Phase C has no D1.5) & False (\texttt{d15\_used}\linebreak[1]\texttt{\_for\_outcome} = false) & D1.5 never used as outcome basis. \\
Conservative conclusion & \texttt{P\_E\_R} & \texttt{P\_E\_R} & Consistent partial external-replication across two external Replogle datasets; NOT a STRONG cross-dataset claim. \\
\bottomrule
\end{tabularx}
\end{table}

%% file: appendix/appendix_t1_per_seed_metrics.tex
\begin{landscape}
\begin{footnotesize}
\begin{longtable}{@{}lllrrrrrr@{}}
\caption{Full per-seed metrics for each (model, seed) pair in both phases, shown in full for the Geneformer arm as the primary worked example. Path-resolved source artifacts are reported in Appendix Table~\ref{tab:app_t3_artifact_lineage}. Per-seed metrics at this resolution are available for the Geneformer arm because its final-test run persisted seed-level predictions; the scGPT and UCE final-test runners persisted only summary metrics, so those arms are summarized by mean test $R^2$ and seed-gate counts in Table~\ref{tab:signed_gb} and Appendix Table~\ref{tab:app_t4_scgpt_uce_family}.}\label{tab:app_t1_per_seed_metrics} \\
\toprule
phase & dataset label & model name & seed & train r2 & val r2 & test r2 & test spearman & test mae \\
\midrule
\endfirsthead
\multicolumn{9}{l}{\itshape (continued)}\\
\toprule
phase & dataset label & model name & seed & train r2 & val r2 & test r2 & test spearman & test mae \\
\midrule
\endhead
\bottomrule
\endfoot
Phase C & RPE1 (Replogle 2022) & C-MAG-only & 20260527 & 0.717265 & 0.717356 & 0.726907 & 0.895341 & 0.286525 \\
Phase C & RPE1 (Replogle 2022) & C-MAG-only & 20260528 & 0.717265 & 0.717356 & 0.726907 & 0.895341 & 0.286525 \\
Phase C & RPE1 (Replogle 2022) & C-MAG-only & 20260529 & 0.717265 & 0.717356 & 0.726907 & 0.895341 & 0.286525 \\
Phase C & RPE1 (Replogle 2022) & C-MAG-only & 20260530 & 0.717265 & 0.717356 & 0.726907 & 0.895341 & 0.286525 \\
Phase C & RPE1 (Replogle 2022) & C-MAG-only & 20260531 & 0.717265 & 0.717356 & 0.726907 & 0.895341 & 0.286525 \\
Phase C & RPE1 (Replogle 2022) & C-Baseline & 20260527 & 0.989824 & 0.867114 & 0.842132 & 0.941710 & 0.199299 \\
Phase C & RPE1 (Replogle 2022) & C-Baseline & 20260528 & 0.987456 & 0.862740 & 0.853948 & 0.946135 & 0.191978 \\
Phase C & RPE1 (Replogle 2022) & C-Baseline & 20260529 & 0.977571 & 0.857244 & 0.827678 & 0.933045 & 0.205785 \\
Phase C & RPE1 (Replogle 2022) & C-Baseline & 20260530 & 0.961071 & 0.820222 & 0.790610 & 0.938106 & 0.249315 \\
Phase C & RPE1 (Replogle 2022) & C-Baseline & 20260531 & 0.977171 & 0.864378 & 0.840781 & 0.940851 & 0.193542 \\
Phase C & RPE1 (Replogle 2022) & C-Geneformer & 20260527 & 0.950843 & 0.878609 & 0.836484 & 0.942485 & 0.200704 \\
Phase C & RPE1 (Replogle 2022) & C-Geneformer & 20260528 & 0.925861 & 0.856495 & 0.817505 & 0.943288 & 0.231086 \\
Phase C & RPE1 (Replogle 2022) & C-Geneformer & 20260529 & 0.945927 & 0.871258 & 0.838315 & 0.943826 & 0.209890 \\
Phase C & RPE1 (Replogle 2022) & C-Geneformer & 20260530 & 0.946533 & 0.882360 & 0.849256 & 0.947415 & 0.193464 \\
Phase C & RPE1 (Replogle 2022) & C-Geneformer & 20260531 & 0.948887 & 0.878832 & 0.839216 & 0.940854 & 0.200590 \\
Phase C & RPE1 (Replogle 2022) & C-Gaussian & 20260527 & 0.996749 & 0.642301 & 0.674623 & 0.862131 & 0.310352 \\
Phase C & RPE1 (Replogle 2022) & C-Gaussian & 20260528 & 0.996151 & 0.675960 & 0.654171 & 0.840158 & 0.329634 \\
Phase C & RPE1 (Replogle 2022) & C-Gaussian & 20260529 & 0.996567 & 0.658084 & 0.638246 & 0.831210 & 0.332607 \\
Phase C & RPE1 (Replogle 2022) & C-Gaussian & 20260530 & 0.995473 & 0.645793 & 0.665053 & 0.846368 & 0.319767 \\
Phase C & RPE1 (Replogle 2022) & C-Gaussian & 20260531 & 0.994816 & 0.632240 & 0.661277 & 0.838688 & 0.318370 \\
Phase C & RPE1 (Replogle 2022) & C-Shuffle & 20260527 & 0.904240 & 0.636413 & 0.657355 & 0.851154 & 0.321101 \\
Phase C & RPE1 (Replogle 2022) & C-Shuffle & 20260528 & 0.868682 & 0.598005 & 0.639418 & 0.835303 & 0.340895 \\
Phase C & RPE1 (Replogle 2022) & C-Shuffle & 20260529 & 0.891056 & 0.630120 & 0.642387 & 0.823899 & 0.345976 \\
Phase C & RPE1 (Replogle 2022) & C-Shuffle & 20260530 & 0.888199 & 0.639807 & 0.632136 & 0.822011 & 0.338921 \\
Phase C & RPE1 (Replogle 2022) & C-Shuffle & 20260531 & 0.883367 & 0.661436 & 0.649414 & 0.839878 & 0.329982 \\
Phase D & K562 GWPS (Replogle 2022) & D-MAG-only & 20260527 & 0.751368 & 0.783030 & 0.577779 & 0.735050 & 0.130472 \\
Phase D & K562 GWPS (Replogle 2022) & D-MAG-only & 20260528 & 0.751368 & 0.783030 & 0.577779 & 0.735050 & 0.130472 \\
Phase D & K562 GWPS (Replogle 2022) & D-MAG-only & 20260529 & 0.751368 & 0.783030 & 0.577779 & 0.735050 & 0.130472 \\
Phase D & K562 GWPS (Replogle 2022) & D-MAG-only & 20260530 & 0.751368 & 0.783030 & 0.577779 & 0.735050 & 0.130472 \\
Phase D & K562 GWPS (Replogle 2022) & D-MAG-only & 20260531 & 0.751368 & 0.783030 & 0.577779 & 0.735050 & 0.130472 \\
Phase D & K562 GWPS (Replogle 2022) & D-Baseline & 20260527 & 0.982324 & 0.884240 & 0.760194 & 0.821858 & 0.094860 \\
Phase D & K562 GWPS (Replogle 2022) & D-Baseline & 20260528 & 0.979745 & 0.881042 & 0.771082 & 0.799419 & 0.095948 \\
Phase D & K562 GWPS (Replogle 2022) & D-Baseline & 20260529 & 0.971827 & 0.875564 & 0.756518 & 0.802427 & 0.095716 \\
Phase D & K562 GWPS (Replogle 2022) & D-Baseline & 20260530 & 0.981459 & 0.880215 & 0.763717 & 0.811370 & 0.096400 \\
Phase D & K562 GWPS (Replogle 2022) & D-Baseline & 20260531 & 0.981738 & 0.882103 & 0.774431 & 0.809929 & 0.095420 \\
Phase D & K562 GWPS (Replogle 2022) & D-Geneformer & 20260527 & 0.941218 & 0.905172 & 0.804144 & 0.910973 & 0.076033 \\
Phase D & K562 GWPS (Replogle 2022) & D-Geneformer & 20260528 & 0.936139 & 0.907908 & 0.788591 & 0.907065 & 0.077731 \\
Phase D & K562 GWPS (Replogle 2022) & D-Geneformer & 20260529 & 0.923598 & 0.897232 & 0.790872 & 0.901563 & 0.078086 \\
Phase D & K562 GWPS (Replogle 2022) & D-Geneformer & 20260530 & 0.926641 & 0.892390 & 0.799728 & 0.907524 & 0.080025 \\
Phase D & K562 GWPS (Replogle 2022) & D-Geneformer & 20260531 & 0.925283 & 0.892762 & 0.788524 & 0.911711 & 0.079244 \\
Phase D & K562 GWPS (Replogle 2022) & D-Gaussian & 20260527 & 0.986257 & 0.734538 & 0.549410 & 0.647376 & 0.139543 \\
Phase D & K562 GWPS (Replogle 2022) & D-Gaussian & 20260528 & 0.987149 & 0.744045 & 0.533531 & 0.596437 & 0.146497 \\
Phase D & K562 GWPS (Replogle 2022) & D-Gaussian & 20260529 & 0.985629 & 0.730645 & 0.532466 & 0.621561 & 0.142063 \\
Phase D & K562 GWPS (Replogle 2022) & D-Gaussian & 20260530 & 0.985607 & 0.738560 & 0.545083 & 0.610404 & 0.143866 \\
Phase D & K562 GWPS (Replogle 2022) & D-Gaussian & 20260531 & 0.984923 & 0.732753 & 0.550697 & 0.615256 & 0.138787 \\
Phase D & K562 GWPS (Replogle 2022) & D-Shuffle & 20260527 & 0.861556 & 0.737535 & 0.533318 & 0.623273 & 0.140473 \\
Phase D & K562 GWPS (Replogle 2022) & D-Shuffle & 20260528 & 0.837479 & 0.745881 & 0.538123 & 0.626253 & 0.141171 \\
Phase D & K562 GWPS (Replogle 2022) & D-Shuffle & 20260529 & 0.845771 & 0.733127 & 0.526541 & 0.597590 & 0.142083 \\
Phase D & K562 GWPS (Replogle 2022) & D-Shuffle & 20260530 & 0.851006 & 0.749403 & 0.533148 & 0.625689 & 0.141292 \\
Phase D & K562 GWPS (Replogle 2022) & D-Shuffle & 20260531 & 0.869138 & 0.735043 & 0.526214 & 0.624639 & 0.142200 \\
\end{longtable}
\end{footnotesize}
\end{landscape}

%% file: appendix/appendix_t2_per_seed_deltas.tex
\begin{landscape}
\begin{footnotesize}
\setlength{\tabcolsep}{6pt}
\begin{longtable}{@{}llrrcrc@{}}
\caption{Per-seed paired deltas with decision-threshold pass flags, shown in full for the Geneformer arm as the primary worked example. ``$\geq$ +0.020'' and ``$\geq$ +0.050'' columns mark whether each per-seed delta met the STRONG seed-consistency threshold for that comparator. Path-resolved source artifacts are reported in Appendix Table~\ref{tab:app_t3_artifact_lineage}. Per-seed deltas at this resolution are available for the Geneformer arm because its final-test run persisted seed-level predictions; the scGPT and UCE final-test runners persisted only summary metrics, so those arms are summarized by their signed baseline-relative deltas and seed-gate counts in Table~\ref{tab:signed_gb}.}\label{tab:app_t2_per_seed_deltas} \\
\toprule
Phase & Dataset & Seed & GF $-$ Baseline $\Delta$ & $\geq$ +0.020 & GF $-$ Gaussian $\Delta$ & $\geq$ +0.050 \\
\midrule
\endfirsthead
\multicolumn{7}{l}{\itshape (continued)}\\
\toprule
Phase & Dataset & Seed & GF $-$ Baseline $\Delta$ & $\geq$ +0.020 & GF $-$ Gaussian $\Delta$ & $\geq$ +0.050 \\
\midrule
\endhead
\bottomrule
\endfoot
Phase C & RPE1 (Replogle 2022)       & 20260527 & $-$0.005648 & False & 0.161861 & True \\
Phase C & RPE1 (Replogle 2022)       & 20260528 & $-$0.036442 & False & 0.163334 & True \\
Phase C & RPE1 (Replogle 2022)       & 20260529 &    0.010637 & False & 0.200068 & True \\
Phase C & RPE1 (Replogle 2022)       & 20260530 &    0.058646 & True  & 0.184203 & True \\
Phase C & RPE1 (Replogle 2022)       & 20260531 & $-$0.001564 & False & 0.177939 & True \\
\midrule
Phase D & K562 GWPS (Replogle 2022)  & 20260527 & 0.043950 & True  & 0.254733 & True \\
Phase D & K562 GWPS (Replogle 2022)  & 20260528 & 0.017509 & False & 0.255060 & True \\
Phase D & K562 GWPS (Replogle 2022)  & 20260529 & 0.034354 & True  & 0.258406 & True \\
Phase D & K562 GWPS (Replogle 2022)  & 20260530 & 0.036011 & True  & 0.254645 & True \\
Phase D & K562 GWPS (Replogle 2022)  & 20260531 & 0.014093 & False & 0.237828 & True \\
\end{longtable}
\end{footnotesize}
\end{landscape}

%% file: appendix/appendix_t3_artifact_lineage.tex
\begin{landscape}
\begin{footnotesize}
\setlength{\extrarowheight}{2pt}
\setlength{\tabcolsep}{4pt}
\begin{longtable}{@{}p{2.2cm}p{1.6cm}p{3.6cm}p{6.5cm}p{1.8cm}p{4.5cm}@{}}
\caption{Artifact lineage and commit history for the full three-model, both-phase evaluation pipeline. The Geneformer arm is shown for Phase C (RPE1) and Phase D (K562 GWPS); the scGPT and UCE arms are shown for their committed final-test and extraction families. Long paths are typeset with break-friendly sequence splitting. Commit hashes, messages, and one primary key-output path per stage are read verbatim from the committed git history. Phase C data-array bytes (labels, features, unit-delta arrays) are gitignored; the key-output cells for those stages therefore cite the committed manifest or record files rather than the arrays. The scGPT and UCE final-test outcomes are additionally locked in committed outcome-record QC files. No separate Phase C interpretation-memo stage exists; the Geneformer Phase C lineage ends at C3-B2.}\label{tab:app_t3_artifact_lineage} \\
\toprule
artifact stage & commit hash & commit message & key outputs & frozen & source artifacts \\
\midrule
\endfirsthead
\multicolumn{6}{l}{\itshape (continued)}\\
\toprule
artifact stage & commit hash & commit message & key outputs & frozen & source artifacts \\
\midrule
\endhead
\bottomrule
\endfoot
\multicolumn{6}{@{}l}{\textbf{Geneformer arm, Phase C (RPE1)}}\\
\midrule
Phase C0-A & \texttt{35cd9f6} & Add Phase C0-A feasibility audits for RPE1 external validation & \texttt{\seqsplit{data/metadata/phasec\_external\_dataset\_feasibility\_replogle\_v1.md}} & yes & git log \\
Phase C pre-registration & \texttt{06482fe} & Pre-register Phase C RPE1 external validation v1 & \texttt{\seqsplit{data/metadata/phasec\_rpe1\_external\_validation\_preregistration\_v1.md}} & yes & git log \\
Phase C1 & \texttt{0d64eca} & Add Phase C1 RPE1 AD-label generation outputs & \texttt{\seqsplit{data/metadata/phasec\_rpe1\_ad\_label\_generation\_manifest\_v1.json}} & yes & git log \\
Phase C2-A & \texttt{2c012f9} & Add Phase C2-A RPE1 feature-construction smoke test & \texttt{\seqsplit{data/metadata/phasec\_rpe1\_feature\_construction\_smoke\_test\_manifest\_v1.json}} & no & git log \\
Phase C2-B planning audit & \texttt{792ffb7} & Add Phase C2-B full feature extraction planning audit & \texttt{\seqsplit{data/metadata/phasec\_rpe1\_full\_feature\_extraction\_planning\_audit\_manifest\_v1.json}} & no & git log \\
Phase C2-B & \texttt{d0b4901} & Add Phase C2-B full RPE1 feature extraction outputs & \texttt{\seqsplit{data/metadata/phasec\_rpe1\_full\_feature\_extraction\_manifest\_v1.json}} & yes & git log \\
Phase C3-A & \texttt{1886803} & Add Phase C3-A RPE1 pre-flight audit and design lock & \texttt{\seqsplit{data/metadata/phasec\_rpe1\_c3a\_preflight\_audit\_manifest\_v1.json}} & yes & git log \\
Phase C3-B1 & \texttt{c3f70b5} & Add Phase C3-B1 train-val sanity execution & \texttt{\seqsplit{results/phasec\_rpe1\_c3b1\_train\_val\_sanity/train\_val\_metrics\_summary.csv}} & yes & git log \\
Phase C3-B2 & \texttt{f72f289} & Add Phase C3-B2 final test evaluation & \texttt{\seqsplit{results/phasec\_rpe1\_c3b2\_final\_test\_evaluation/final\_test\_outcome.json}} & yes & git log \\
\midrule
\multicolumn{6}{@{}l}{\textbf{Geneformer arm, Phase D (K562 GWPS)}}\\
\midrule
Phase D0-A & \texttt{278b5e1} & Add Phase D0-A K562 GWPS feasibility audit & \texttt{\seqsplit{data/metadata/phase\_d\_big\_dataset\_d0a\_feasibility\_manifest\_v1.json}} & yes & git log \\
Phase D0-B1 & \texttt{98f5779} & Add Phase D0-B1 design decision review & \texttt{\seqsplit{data/metadata/phase\_d\_big\_dataset\_d0b1\_design\_decision\_manifest\_v1.json}} & yes & git log \\
Phase D0-B2 & \texttt{670f783} & Pre-register Phase D K562 GWPS external validation & \texttt{\seqsplit{data/metadata/phase\_d\_big\_dataset\_d0b2\_preregistration\_manifest\_v1.json}} & yes & git log \\
Phase D0-B2 clarification & \texttt{0bbe161} & Clarify Phase D0-B2 MAG x\_delta preprocessing before full extraction & \texttt{\seqsplit{data/metadata/phase\_d\_big\_dataset\_d0b2\_mag\_xdelta\_preprocessing\_clarification\_manifest\_v1.json}} & yes & git log \\
Phase D1 & \texttt{23214b1} & Add Phase D1 K562 GWPS primary label construction & \texttt{\seqsplit{data/processed/phase\_d\_big\_dataset\_d1\_primary\_labels/primary\_global\_nt\_ad\_labels.parquet}} & yes & git log \\
Phase D1.5 & \texttt{db2c83a} & Add Phase D1.5 K562 GWPS secondary per-gem\_group diagnostic label & \texttt{\seqsplit{data/processed/phase\_d\_big\_dataset\_d15\_per\_gem\_group\_diagnostic/secondary\_per\_gem\_group\_ad\_labels.parquet}} (diagnostic only) & yes & git log \\
Phase D2-A & \texttt{e23e481} & Add Phase D2-A K562 GWPS feature construction smoke test & \texttt{\seqsplit{results/phase\_d\_big\_dataset\_d2a\_feature\_smoke/d2a\_feature\_smoke\_qc.json}} & yes & git log \\
Phase D2-B & \texttt{980473d} & Add Phase D2-B K562 GWPS full feature construction outputs & \texttt{\seqsplit{results/phase\_d\_big\_dataset\_d2b\_full\_features/d2b\_full\_feature\_qc.json}} & yes & git log \\
Phase D3-A & \texttt{e46a019} & Add Phase D3-A K562 GWPS pre-flight audit and design lock & \texttt{\seqsplit{results/phase\_d\_big\_dataset\_d3a\_preflight\_audit/d3a\_preflight\_audit\_qc.json}} & yes & git log \\
Phase D3-B1 & \texttt{8d1a5c6} & Add Phase D3-B1 K562 GWPS train-validation sanity outputs & \texttt{\seqsplit{results/phase\_d\_big\_dataset\_d3b1\_train\_val\_sanity/d3b1\_train\_val\_qc.json}} & yes & git log \\
Phase D3-B2 & \texttt{2a4e6a3} & Add Phase D3-B2 K562 GWPS final test once outputs & \texttt{\seqsplit{results/phase\_d\_big\_dataset\_d3b2\_final\_test\_once/d3b2\_final\_outcome\_decision.json}} & yes & git log \\
Phase D3-C & \texttt{1b2eff1} & Add Phase D3-C K562 GWPS final interpretation memo and paper-ready outputs & \texttt{\seqsplit{results/phase\_d\_big\_dataset\_d3c\_final\_interpretation/d3c\_key\_claims.json}} & yes & git log \\
\midrule
\multicolumn{6}{@{}l}{\textbf{scGPT arm, final-test-once (P2-B4)}}\\
\midrule
P2-B4 scGPT final-test, Phase C & \texttt{e5b53ab} & Add P2-B4 scGPT final-test-once evaluation & \texttt{\seqsplit{paper/mada\_geneformer\_p2b4\_scgpt\_final\_test\_once\_v1/p2b4\_phase\_c\_final\_test\_metrics\_summary.csv}} & yes & git log \\
P2-B4 scGPT final-test, Phase D & \texttt{e5b53ab} & Add P2-B4 scGPT final-test-once evaluation & \texttt{\seqsplit{paper/mada\_geneformer\_p2b4\_scgpt\_final\_test\_once\_v1/p2b4\_phase\_d\_final\_test\_metrics\_summary.csv}} & yes & git log \\
\midrule
\multicolumn{6}{@{}l}{\textbf{UCE arm, full extraction and final-test-once (P2-D3.5B)}}\\
\midrule
UCE Phase C row-level construction and checkpoint & \texttt{9170ee7} & Add P2-D3.5B Phase C row-level UCE construction and checkpoint & \texttt{\seqsplit{paper/mada\_geneformer\_p2d35b\_uce\_full\_extraction\_v1/p2d35b\_phase\_c\_checkpoint\_v1.md}} & yes & git log \\
UCE Phase D row-level construction and checkpoint & \texttt{5ab90ea} & Add P2-D3.5B Phase D row-level UCE construction and checkpoint & \texttt{\seqsplit{paper/mada\_geneformer\_p2d35b\_uce\_full\_extraction\_v1/p2d35b\_phase\_d\_checkpoint\_v1.md}} & yes & git log \\
UCE comparability audit (Phase C + D) & \texttt{db23bb7} & Add P2-D3.5B UCE comparability audit & \texttt{\seqsplit{paper/mada\_geneformer\_p2d35b\_uce\_full\_extraction\_v1/p2d35b\_uce\_comparability\_audit\_v1.md}} & yes & git log \\
UCE train-val preflight, Phase C & \texttt{eb6f834} & Add P2-D3.5B UCE train-val preflight & \texttt{\seqsplit{paper/mada\_geneformer\_p2d35b\_uce\_full\_extraction\_v1/p2d35b\_phase\_c\_uce\_train\_val\_metrics\_summary.csv}} & yes & git log \\
UCE train-val preflight, Phase D & \texttt{eb6f834} & Add P2-D3.5B UCE train-val preflight & \texttt{\seqsplit{paper/mada\_geneformer\_p2d35b\_uce\_full\_extraction\_v1/p2d35b\_phase\_d\_uce\_train\_val\_metrics\_summary.csv}} & yes & git log \\
UCE final-test, Phase C & \texttt{d1c263f} & Add P2-D3.5B UCE final-test once & \texttt{\seqsplit{paper/mada\_geneformer\_p2d35b\_uce\_full\_extraction\_v1/p2d35b\_phase\_c\_uce\_final\_test\_metrics\_summary.csv}} & yes & git log \\
UCE final-test, Phase D & \texttt{d1c263f} & Add P2-D3.5B UCE final-test once & \texttt{\seqsplit{paper/mada\_geneformer\_p2d35b\_uce\_full\_extraction\_v1/p2d35b\_phase\_d\_uce\_final\_test\_metrics\_summary.csv}} & yes & git log \\
\end{longtable}

\smallskip
\noindent\textit{Frozen-status provenance (Geneformer Phase C).} The frozen verdicts for the Phase C stages were determined by reading the committed stage manifests and validation records, not inferred from commit titles. Six stages carry explicit committed lock or pass evidence (for example, the pre-registration record states it ``locks the Phase C protocol''; the C3-A manifest records a \texttt{design lock} with \texttt{C3A\_PREFLIGHT\_PASS}; and C1, C2-B full extraction, and C3-B1 record locked-split, \texttt{PASS}, and \texttt{c3b\_final\_evaluation\_run = false} status respectively). Two stages are marked \texttt{no} because their own committed manifests document them as non-final rather than frozen: Phase C2-A outputs are recorded as \texttt{SMOKE\_TEST\_NT\_SUBSET\_ONLY\_NOT\_FINAL} with a \texttt{NEEDS\_REVIEW} decision, and the Phase C2-B planning audit records \texttt{C2B\_PLANNING\_NEEDS\_REVIEW}. These two entries reflect documented non-frozen status, not missing evidence.
\end{footnotesize}
\end{landscape}

%% file: tables/app_t4_scgpt_uce_family.tex
\begin{table}[t]
\caption{Final-test performance (mean test $R^2$ across the five locked
training seeds) for the full model families of the scGPT and UCE arms, in
Phase C and Phase D. These two arms share the same committed Baseline and
MAG-only branch within their chain (separately fitted from, and differing
slightly from, the Geneformer arm's); the Geneformer arm's corresponding
family is reported in main-text Table~\ref{tab:t2_final_performance}. The
signed baseline-relative deltas for all three arms are given in
Table~\ref{tab:signed_gb}.}
\label{tab:app_t4_scgpt_uce_family}
\centering
\small
\begin{tabular}{@{}lllr@{}}
\toprule
Arm & Phase & Branch & mean test $R^2$ \\
\midrule
scGPT & C (RPE1) & Baseline          & 0.8343 \\
scGPT & C (RPE1) & scGPT (primary)   & 0.8156 \\
scGPT & C (RPE1) & Gaussian (512-d)  & 0.6597 \\
scGPT & C (RPE1) & Shuffle           & 0.6700 \\
scGPT & C (RPE1) & MAG-only          & 0.7269 \\
\addlinespace[2pt]
scGPT & D (K562) & Baseline          & 0.7607 \\
scGPT & D (K562) & scGPT (primary)   & 0.7387 \\
scGPT & D (K562) & Gaussian (512-d)  & 0.5225 \\
scGPT & D (K562) & Shuffle           & 0.5750 \\
scGPT & D (K562) & MAG-only          & 0.5778 \\
\midrule
UCE & C (RPE1) & Baseline           & 0.8343 \\
UCE & C (RPE1) & UCE (primary)      & 0.8294 \\
UCE & C (RPE1) & Gaussian (1280-d)  & 0.6488 \\
UCE & C (RPE1) & Shuffle            & 0.6652 \\
UCE & C (RPE1) & MAG-only           & 0.7269 \\
\addlinespace[2pt]
UCE & D (K562) & Baseline           & 0.7607 \\
UCE & D (K562) & UCE (primary)      & 0.6819 \\
UCE & D (K562) & Gaussian (1280-d)  & 0.5326 \\
UCE & D (K562) & Shuffle            & 0.5737 \\
UCE & D (K562) & MAG-only           & 0.5778 \\
\bottomrule
\end{tabular}
\end{table}

%% file: appendix/appendix_a4_allowed_disallowed_claims.tex
\begin{table}[t]
\caption{Allowed and disallowed claims (reviewer-facing wording transparency). The Geneformer-arm entries source from \texttt{d3c\_key\_claims.json}; the cross-architecture entries covering the scGPT and UCE arms are derived from the committed three-arm final-test summaries (Table~\ref{tab:signed_gb} and Appendix Table~\ref{tab:app_t4_scgpt_uce_family}). \emph{Wording constraint:} ``Use literal phrasing; do not soften or strengthen'' for all allowed claims; ``Never appears in Abstract/Results/Discussion'' for all disallowed claims.}
\label{tab:app_a4_allowed_disallowed_claims}
\centering
\footnotesize
\renewcommand{\arraystretch}{1.18}
\setlength{\tabcolsep}{6pt}
\begin{tabularx}{\textwidth}{@{}>{\raggedright\arraybackslash}p{1.8cm} >{\raggedright\arraybackslash}X@{}}
\toprule
Claim type & Claim text \\
\midrule
allowed & Geneformer showed \texttt{PARTIAL\_EXTERNAL\_REPLICATION} in the K562 GWPS external evaluation under the locked D0-B2 decision rule. \\
allowed & D-Geneformer improved mean test $R^2$ over the expression baseline (D-Baseline) by +0.02918 (0.7944 vs 0.7652). \\
allowed & D-Geneformer exceeded the capacity-matched Gaussian control (D-Gaussian) by a mean test $R^2$ of +0.25213 with 5/5 seed consistency at the +0.050 bar. \\
allowed & D-Geneformer was far above the row-identity shuffle control (D-Shuffle); $|$Shuffle $-$ Geneformer$|$ mean test $R^2$ = 0.26290, far from the 0.020 invalidity threshold. \\
allowed & Random capacity alone does NOT explain D-Geneformer's gain: Gaussian $-$ Baseline mean test $R^2$ = $-$0.22295, so the random-capacity branch is actively worse than the expression baseline. \\
allowed & Validation parity vs Phase D3-B1 holds at machine epsilon (max $|\Delta|$ = 1.11e$-$16 across 25 model $\times$ seed pairs); the final-test pipeline is numerically deterministic to machine precision with the train/val sanity pipeline. \\
allowed & The result did NOT meet the locked \texttt{STRONG\_EXTERNAL\_REPLICATION} criterion because the GF $-$ Baseline seed-consistency rule required $\geq$ 4/5 seeds with delta $\geq$ +0.020 and only 3/5 seeds met it. \\
allowed & The result was NOT invalidated by the shuffle control and was NOT capacity-only. \\
allowed & If Phase C RPE1 is also \texttt{PARTIAL\_EXTERNAL\_REPLICATION}, the partial external-replication pattern appears consistent across two external Replogle datasets. \\
allowed & scGPT and UCE each also closed as \texttt{PARTIAL\_EXTERNAL\_REPLICATION} in both external phases under the same locked decision rule, so all three foundation-model arms share the \texttt{PARTIAL\_PARTIAL} cross-dataset pattern. \\
allowed & Neither scGPT nor UCE improved on its own expression baseline: their signed baseline-relative deltas were negative or near zero in both phases (Table~\ref{tab:signed_gb}), while all three arms cleared their matched-capacity Gaussian and shuffle controls. \\
allowed & The shared \texttt{PARTIAL} outcome across three architecturally distinct representations (Geneformer 768-d, scGPT 512-d, UCE 1280-d) indicates a representation-agnostic result at the outcome-category level, not a Geneformer-specific one; the signed baseline-relative delta still differs across arms (Table~\ref{tab:signed_gb}). \\
\midrule
disallowed & Do NOT claim \texttt{STRONG\_EXTERNAL\_REPLICATION}. \\
disallowed & Do NOT claim strong external generalization, confirmed strong replication, or strong cross-dataset replication. \\
disallowed & Do NOT claim Phase C + Phase D together upgrade the outcome to STRONG. \\
disallowed & Do NOT claim the model is production-ready or clinically deployable. \\
disallowed & Do NOT claim the +0.020 or +0.050 threshold should have been lower. \\
disallowed & Do NOT reinterpret the locked outcome after seeing test results. \\
disallowed & Do NOT use D1.5 (secondary per-gem\_group diagnostic) to change the outcome. \\
disallowed & Do NOT claim Geneformer dominates expression-derived features at large effect size. \\
disallowed & Do NOT claim the outcome would have been STRONG without the seed-consistency rule. \\
disallowed & Do NOT claim the framework discriminates between foundation models or ranks them; the separation among the three arms is at the criterion level (signed baseline-relative delta), not a locked-outcome difference. \\
disallowed & Do NOT claim scGPT or UCE beat the expression baseline; both were at or below their own baselines in both phases. \\
disallowed & Do NOT claim per-seed or confidence-interval resolution for the scGPT or UCE arms; their final-test runners persisted only summary metrics, not row-level predictions. \\
\bottomrule
\end{tabularx}
\end{table}

%% file: appendix/appendix_t5_repro_checklist.tex
{\footnotesize
\setlength{\tabcolsep}{3pt}
\renewcommand{\arraystretch}{1.15}
\setlength{\LTleft}{0pt}
\setlength{\LTright}{0pt}

\begin{longtable}{@{}>{\raggedright\arraybackslash}p{3.55cm}>{\raggedright\arraybackslash}p{1.45cm}>{\raggedright\arraybackslash}p{8.55cm}@{}}
\caption{Reviewer-facing reproducibility checklist covering all three foundation-model arms (Geneformer, scGPT, UCE) across both external phases. SHA-256 fingerprints are reported only when an exact committed 64-character value is available. Entries marked ``n/a (see lineage)'' indicate that no separate arm-specific committed field exists for that item; corresponding lineage information is reported in Appendix Table~\ref{tab:app_t3_artifact_lineage}.}
\label{tab:app_t5_repro_checklist} \\
\toprule
Item & Status & Value (and source) \\
\midrule
\endfirsthead

\multicolumn{3}{l}{\textit{Reproducibility checklist (continued)}}\\
\toprule
Item & Status & Value (and source) \\
\midrule
\endhead

\bottomrule
\endfoot

\multicolumn{3}{@{}l}{\textbf{Geneformer arm, Phase D (K562 GWPS)}}\\
\midrule
no training rerun & PASS & true \\
no test rerun & PASS & true \\
no new predictions & PASS & true \\
no threshold change & PASS & true \\
no outcome change & PASS & true \\
final-test-once & PASS & true (\texttt{\seqsplit{test\_evaluation\_run=true}}; outcome decision made once) \\
validation parity (D3-B2 vs D3-B1) & PASS & max $|\Delta R^2| = 1.110\mathrm{e}{-16}$; max $|\Delta \rho| = 1.110\mathrm{e}{-16}$; max $|\Delta \mathrm{MAE}| = 8.327\mathrm{e}{-17}$; tol $= 1\mathrm{e}{-6}$ \\
D1 primary-label sha256 verified & PASS & \texttt{\seqsplit{79d5453bc1854c32019469e9cf9642920615effce85340d3c1db82d60170b342}} \\
D3-A split assignment sha256 verified & PASS & \texttt{\seqsplit{31fd1c6b032557c82d88f7664c34cd0627a89eff598ff9ad99ff17b2ce2f39e1}} \\
\texttt{target\_gene} leakage count = 0 & PASS & 0 \\
D1.5 used for outcome & PASS (false) & false \\
Co-Authored-By count in D3-B2 commit & PASS & 0 (\texttt{2a4e6a3}) \\
Co-Authored-By count in D3-C commit & PASS & 0 (\texttt{1b2eff1}) \\

\midrule
\multicolumn{3}{@{}l}{\textbf{Geneformer arm, Phase C (RPE1)}}\\
\midrule
final-test-once & PASS & true (\texttt{\seqsplit{test\_evaluated\_once=true}}; \texttt{\seqsplit{no\_model\_tuning\_after\_test=true}}; \texttt{\seqsplit{no\_design\_change\_after\_test=true}}) \\
validation parity (C3-B2 vs C3-B1) & PASS & aggregate max $|\Delta| = 1.110\mathrm{e}{-16}$; per-row $|\Delta R^2| = 0.0$ for all 25 (model, seed) pairs; tol $= 1\mathrm{e}{-5}$ \\
C1 primary-label sha256 verified & PASS & \texttt{\seqsplit{200675e282418b4af091f503190a56d72339cb629989b5467e96e608764224a6}} (cross-checked in the C3-A manifest) \\
C3-A split-assignment sha256 & n/a (see lineage) & no dedicated Phase C split-assignment parquet sha256 is committed; the locked \texttt{target\_gene}-grouped split (seed 20260601) is documented through split-integrity counts \\
\texttt{target\_gene} leakage count = 0 & PASS & 0 (\texttt{\seqsplit{genes\_in\_multiple\_splits=0}}; \texttt{\seqsplit{gene\_transcript\_in\_multiple\_splits=0}}) \\
Co-Authored-By count in C3-B2 commit & PASS & 0 (\texttt{f72f289}) \\

\midrule
\multicolumn{3}{@{}l}{\textbf{scGPT arm, final-test-once (P2-B4)}}\\
\midrule
final-test-once (both phases) & PASS & true (\texttt{\seqsplit{no\_test\_rerun}}; \texttt{\seqsplit{no\_lock\_change}}; \texttt{\seqsplit{no\_extraction\_rerun}}; \texttt{\seqsplit{test\_touched\_exactly\_once}} for C and D) \\
validation parity (C; D) & PASS & Phase C max $|\Delta| = 1.110\mathrm{e}{-16}$ (tol $1\mathrm{e}{-5}$); Phase D max $|\Delta| = 8.327\mathrm{e}{-17}$ (tol $1\mathrm{e}{-6}$) \\
arm-specific sha256 fingerprint & n/a (see lineage) & no committed scGPT-specific output sha256 is recorded in the P2-B4 QC/manifest files; the arm reuses the locked Phase C label parquet \texttt{\seqsplit{200675e282418b4af091f503190a56d72339cb629989b5467e96e608764224a6}} \\
\texttt{target\_gene} leakage evidence & PARTIAL & Phase C carries a committed runtime split-leakage assertion requiring zero \texttt{target\_gene} groups spanning more than one split; a separate committed exact leakage count for the scGPT arm is \texttt{NOT FOUND}; source: \texttt{\seqsplit{paper/mada\_geneformer\_p2b4\_scgpt\_final\_test\_once\_v1/\_p2b4\_final\_test\_runner.py}} \\
locked outcome (C; D) & FROZEN & \texttt{PARTIAL\_EXTERNAL\_REPLICATION}; \texttt{PARTIAL\_EXTERNAL\_REPLICATION} \\
Co-Authored-By count (commit e5b53ab) & PASS & 0 (\texttt{e5b53ab}) \\

\midrule
\multicolumn{3}{@{}l}{\textbf{UCE arm, full extraction and final-test-once (P2-D3.5B)}}\\
\midrule
final-test-once (both phases) & PASS & true (\texttt{\seqsplit{uce\_final\_test\_once\_completed=true}}; \texttt{\seqsplit{rerun\_permitted=false}}; \texttt{\seqsplit{no\_final\_test\_rerun\_in\_this\_continuation=true}}) \\
validation parity (C; D) & PASS & Phase C max $|\Delta| = 1.110\mathrm{e}{-16}$ (tol $1\mathrm{e}{-5}$); Phase D max $|\Delta| = 8.327\mathrm{e}{-17}$ (tol $1\mathrm{e}{-6}$) \\
Phase C row-level $z$-unit-delta sha256 & PASS & \texttt{\seqsplit{abc76c49f2f587a992c56cf45ea622cdd73cd580d89ef3c8f312bf631dd2a545}} \\
Phase D row-level $z$-unit-delta sha256 & PASS & \texttt{\seqsplit{04b734ab2b12bb0bc3090ccf6dccd1147bc5aafe30400158bd5f2fda060f8d2a}} \\
extraction bridge sha256 match & PASS & true (\texttt{\seqsplit{fabfa0c2f49c598c59ae432a32c3499a5908c033756c663b5e0cddf58deea8e1}}) \\
\texttt{target\_gene} leakage & PASS & no test row appears in any train or val slice; enforced by a runtime split-leakage assertion \\
locked outcome (C; D) & FROZEN & \texttt{PARTIAL\_EXTERNAL\_REPLICATION}; \texttt{PARTIAL\_EXTERNAL\_REPLICATION} \\
Co-Authored-By count (commits 9170ee7 / 5ab90ea / eb6f834 / d1c263f) & PASS & 0 (each) \\

\midrule
\multicolumn{3}{@{}l}{\textbf{Cross-arm summary}}\\
\midrule
Phase D outcome (all three arms) & FROZEN & \texttt{PARTIAL\_EXTERNAL\_REPLICATION} \\
Phase C outcome (all three arms) & FROZEN & \texttt{PARTIAL\_EXTERNAL\_REPLICATION} \\
cross-dataset pattern (all three arms) & PASS & \texttt{PARTIAL\_PARTIAL} \\

\end{longtable}
}